\documentclass[twocolumn]{aastex631}
\usepackage{hyperref}
\usepackage{graphicx}
\usepackage{natbib}
\usepackage{float}
\usepackage{wrapfig}
\usepackage{amsmath}
\newcommand{\squishlist}{
 \begin{list}{$\bullet$}
  { \setlength{\itemsep}{1pt}
     \setlength{\parsep}{0pt}
     \setlength{\topsep}{1pt}
     \setlength{\partopsep}{0pt}
     \setlength{\leftmargin}{1.5em}
     \setlength{\labelwidth}{1.5em}
     \setlength{\labelsep}{0.5em} } }
\newcommand{\squishend}{
  \end{list}  }
\usepackage{upgreek} 

\shorttitle{MIRI dayside emission of WASP-17b}
\shortauthors{Valentine et al.}

\begin{document}

\title{JWST-TST DREAMS: Non-Uniform Dayside Emission for WASP-17b from MIRI/LRS}


\author[0000-0002-2643-6836]{Daniel Valentine}
\affiliation{University of Bristol, HH Wills Physics Laboratory, Tyndall Avenue, Bristol, UK}
\correspondingauthor{Daniel Valentine}
\email{daniel.valentine@bristol.ac.uk}

\author[0000-0003-4328-3867]{Hannah R. Wakeford}
\affiliation{University of Bristol, HH Wills Physics Laboratory, Tyndall Avenue, Bristol, UK}

\author[0000-0002-8211-6538]{Ryan C. Challener}
\affiliation{Department of Astronomy and Carl Sagan Institute, Cornell University, 122 Sciences Drive, Ithaca, NY 14853, USA}

\author[0000-0003-1240-6844]{Natasha E. Batalha}
\affiliation{NASA Ames Research Center, MS 245-3, Moffett Field, CA 94035, USA}

\author[0000-0002-8507-1304]{Nikole K. Lewis}
\affiliation{Department of Astronomy and Carl Sagan Institute, Cornell University, 122 Sciences Drive, Ithaca, NY 14853, USA}

\author[0000-0001-5878-618X]{David Grant}
\affiliation{University of Bristol, HH Wills Physics Laboratory, Tyndall Avenue, Bristol, UK}

\author[0000-0003-0814-7923]{Elijah Mullens}
\affiliation{Department of Astronomy and Carl Sagan Institute, Cornell University, 122 Sciences Drive, Ithaca, NY 14853, USA}

\author[0000-0001-8703-7751]{Lili Alderson}
\affiliation{University of Bristol, HH Wills Physics Laboratory, Tyndall Avenue, Bristol, UK}

\author[0000-0002-8515-7204]{Jayesh Goyal}
\affiliation{School of Earth and Planetary Sciences (SEPS), National Institute of Science Education and Research (NISER), HBNI, Odisha, India}

\author[0000-0003-4816-3469]{Ryan J. MacDonald}
\affiliation{Department of Astronomy, University of Michigan, 1085 S. University Ave., Ann Arbor, MI 48109, USA}

\author[0000-0002-2739-1465]{Erin M. May}
\affiliation{Johns Hopkins University Applied Physics Laboratory, 11100 Johns Hopkins Rd, Laurel, MD 20723, USA}

\author[0000-0002-6892-6948]{Sara Seager}
\affil{Department of Earth, Atmospheric, and Planetary Sciences, Massachusetts Institute of Technology, Cambridge, MA 02139, USA}
\affiliation{Kavli Institute for Astrophysics and Space Research, Massachusetts Institute of Technology, Cambridge, MA 02139, USA}
\affil{Department of Aeronautics and Astronautics, Massachusetts Institute of Technology, Cambridge, MA 02139, USA}

\author[0000-0002-7352-7941]{Kevin B. Stevenson}
\affiliation{Johns Hopkins University Applied Physics Laboratory, 11100 Johns Hopkins Rd, Laurel, MD 20723, USA}

\author[0000-0003-3305-6281]{Jeff A. Valenti}
\affiliation{Space Telescope Science Institute, 3700 San Martin Drive, Baltimore, MD 21218, USA}

\author[0000-0002-0832-710X]{Natalie H. Allen}
\affiliation{William H. Miller III Department of Physics and Astronomy, Johns Hopkins University, Baltimore, MD 21218, USA}

\author[0000-0001-9513-1449]{N\'{e}stor Espinoza}
\affiliation{Space Telescope Science Institute, 3700 San Martin Drive, Baltimore, MD 21218, USA}
\affiliation{William H. Miller III Department of Physics and Astronomy, Johns Hopkins University, Baltimore, MD 21218, USA}

\author[0000-0002-5322-2315]{Ana Glidden}
\affiliation{Kavli Institute for Astrophysics and Space Research, Massachusetts Institute of Technology, Cambridge, MA 02139, USA}
\affiliation{Department of Earth, Atmospheric and Planetary Sciences, Massachusetts Institute of Technology, Cambridge, MA 02139, USA}

\author[0000-0003-0854-3002]{Am\'{e}lie A. Gressier}
\affiliation{Space Telescope Science Institute, 3700 San Martin Drive, Baltimore, MD 21218, USA}

\author[0000-0001-5732-8531]{Jingcheng Huang}
\affiliation{Department of Earth, Atmospheric and Planetary Sciences, Massachusetts Institute of Technology, Cambridge, MA 02139, USA}

\author[0000-0003-0525-9647]{Zifan Lin}
\affiliation{Department of Earth, Atmospheric and Planetary Sciences, Massachusetts Institute of Technology, Cambridge, MA 02139, USA}

\author[0000-0002-2508-9211]{Douglas Long}
\affiliation{Space Telescope Science Institute, 3700 San Martin Drive, Baltimore, MD 21218, USA}

\author[0000-0002-2457-272X]{Dana R. Louie}
\affiliation{Catholic University of America, Department of Physics, Washington, DC, 20064, USA}
\affiliation{Exoplanets and Stellar Astrophysics Laboratory (Code 667), NASA Goddard Space Flight Center, Greenbelt, MD 20771, USA}
\affiliation{Center for Research and Exploration in Space Science and Technology II, NASA/GSFC, Greenbelt, MD 20771, USA}

\author{Mark Clampin}
\affiliation{NASA Headquarters, 300 E Street SW, Washington, DC 20546, USA}

\author[0000-0002-3191-8151]{Marshall Perrin}
\affiliation{Space Telescope Science Institute, 3700 San Martin Drive, Baltimore, MD 21218, USA}

\author[0000-0001-7827-7825]{Roeland P. van der Marel}
\affiliation{Space Telescope Science Institute, 3700 San Martin Drive, Baltimore, MD 21218, USA}
\affiliation{William H. Miller III Department of Physics and Astronomy, Johns Hopkins University, Baltimore, MD 21218, USA}

\author{C. Matt Mountain}
\affiliation{Association of Universities for Research in Astronomy, 1331 Pennsylvania Avenue NW Suite 1475, Washington, DC 20004, USA}

\begin{abstract}
We present the first spectroscopic characterisation of the dayside atmosphere of WASP-17b in the mid-infrared using a single JWST MIRI/LRS eclipse observation.
From forward-model fits to the 5--12 $\upmu$m emission spectrum, we tightly constrain the heat redistribution factor of WASP-17b to be $0.92\pm0.02$ at the pressures probed by this data, indicative of inefficient global heat redistribution. We also marginally detect a supersolar abundance of water, consistent with previous findings for WASP-17b, but note our weak constraints on this parameter. These results reflect the thermodynamically rich but chemically poor information content of MIRI/LRS emission data for high-temperature hot Jupiters. Using the eclipse mapping method, which utilises the signals that the spatial emission profile of an exoplanet imprints on the eclipse light curve during ingress/egress due to its partial occultation by the host star, we also construct the first eclipse map of WASP-17b, allowing us to diagnose its multidimensional atmospheric dynamics for the first time. We find a day-night temperature contrast of order 1000 K at the pressures probed by this data, consistent with our derived heat redistribution factor, along with an eastward longitudinal hotspot offset of $18.7^{+11.1\circ}_{-3.8}$, indicative of the presence of an equatorial jet induced by day-night thermal forcing being the dominant redistributor of heat from the substellar point. These dynamics are consistent with general circulation model predictions for WASP-17b. This work is part of a series of studies by the JWST Telescope Scientist Team (JWST-TST), in which we use Guaranteed Time Observations to perform Deep Reconnaissance of Exoplanet Atmospheres through Multi-instrument Spectroscopy (DREAMS).
\end{abstract}

\section{Introduction} \label{sec:intro}
WASP-17b is a highly irradiated ($T_{\rm{eq}} = 1770 \, \rm{K}$) hot Jupiter on a 3.735 day retrograde orbit around an F6 type star \citep{anderson2011, triaud2010spin, Alderson2022}. With a radius nearly twice that of Jupiter ($1.932 \, \rm{R}_{\rm{Jup}}$), but less than half the mass ($0.477 \, \rm{M}_{\rm{Jup}}$), WASP-17b has a low density of just $0.067 \, \rm{\rho}_{\rm{Jup}}$ \citep{southworth2012homogeneous}. Its relatively high equilibrium temperature and low density result in a large atmospheric scale height of approximately 2000 km. These factors make this planet ideally suited for multidimensional atmospheric characterisation through both transmission and emission spectroscopy.

Whilst WASP-17b has been characterised extensively in transmission using both ground-based high-resolution instruments \citep{wood2011transmission, zhou2012detection, bento2014optical, sedaghati2016potassium, khalafinejad2018atmosphere} and lower-resolution space-based instruments such as the Hubble Space Telescope \citep[HST;][]{mandell2013exoplanet, sing2016continuum, Alderson2022}, emission measurements have, to date, been lacking. Prior to this program, the only emission measurement of WASP-17b came from two photometric Spitzer points obtained at 4.5 and 8 $\upmu$m \citep{anderson2011}. The JWST Telescope Scientist Team (JWST-TST) Deep Reconnaissance of Exoplanet Atmospheres through Multi-instrument Spectroscopy (DREAMS) program provided the first spectroscopic characterisation of the dayside atmosphere of WASP-17b in the 0.6--2.8 $\upmu$m range using JWST NIRISS/SOSS (A. Gressier et al. accepted, AJ), which yielded firm detections of water, a supersolar metallicity, and excess emission at short wavelengths, the latter possibly attributed to a higher-than-expected internal temperature and/or reflected light from high-albedo clouds. This study by A. Gressier et al. (accepted, AJ) emphasises the fact that exoplanet atmospheres are inherently multidimensional structures, with complex thermal, chemical, and dynamical profiles. Whilst simpler low-dimension models have been adequate in the past to interpret observational data, the quality of JWST light curves necessitate more complex interpretation.

In this study, we observed WASP-17b in eclipse with JWST MIRI/LRS. This instrument has provided a number of comprehensive emission spectra of exoplanet atmospheres in the 5--12 $\upmu$m range \citep[e.g.,][]{gj1214emissionspectrum, wasp43emissionspectrum}. MIRI/LRS is particularly advantageous for determining the thermal structures of hot Jupiters owing to its broad mid-IR coverage, which is the bandpass in which the planetary-to-stellar flux ratio peaks for hot Jupiters in orbit around F-type stars. The highly inflated radius and retrograde orbit of WASP-17b allude to a dynamic and potentially chaotic evolutionary history. Possibilities include interior heating via tidal dissipation to explain its inflated radius \citep{bodenheimer2001}, and planet–planet or star–planet scattering during inward migration to explain its retrograde motion \citep{anderson2011}. This inflation, and hence low gravity, along with the potential for significant internal heating, likely play an important role in shaping the planet’s atmospheric thermal structure and winds. These factors make WASP-17b an excellent target for the type of in-depth multidimensional atmospheric characterisation our program aims to carry out.

Previous modelling work on WASP-17b includes that of \citet{kataria2016}, who conducted an atmospheric circulation study of nine hot Jupiters, including WASP-17b, with comparisons of their general circulation models (GCMs) to Hubble/STIS, Hubble/WFC3, and Spitzer/IRAC data. They showed that across the temperature and rotational period space, hot Jupiters exhibit a wide range of atmospheric dynamics and thermal structures, which produce highly non-uniform brightness patterns on the dayside hemisphere.
These signals are imprinted on the light curve of a transiting exoplanet as it is eclipsed by its host star during ingress/egress due to the varying degree of partial occulation. Using the technique of eclipse mapping, these signals can be extracted and used to recover the planet's dayside photospheric spatial emission profile \citep{rauscher2007}. Due to the geometry at which the stellar limb occults the planet's dayside hemisphere during ingress/egress, eclipse mapping is currently the only method of measuring 2D (longitude-latitude) profiles of exoplanet atmospheres, and at greater-than-hemispheric scale due to the fine spatial sampling. This is advantageous over phase mapping, which exploits the rotation of the planet and can thus only provide meaningful constraints on large-scale longitudinal structure.

Eclipse mapping efforts have been limited in the past by data quality: prior to JWST, the only successful eclipse map was that of HD 189733b using Spitzer/IRAC data \citep{majeau2012, dewit2012}, but even with this most optimal target, eight secondary eclipse measurements and a partial phase curve were required in order to yield a meaningfully constrained map, which itself was low-order, comprising only large-scale bulk features. The superior quality of JWST light curves means that we are now able to produce higher-resolution maps for a larger sample size of exoplanets using fewer datasets. The ultra-hot Jupiter WASP-18b was eclipse mapped from a single NIRISS/SOSS eclipse \citep{wasp18eclipsemap}, producing a map that revealed small-scale structure in the atmosphere and was able to place good constraints on the day-night temperature contrast, longitudinal hotspot location, in addition to giving indications of potential magnetic field effects via Rayleigh drag \citep{beltz2022}.

Two eclipses from a MIRI/LRS phase curve were utilised to successfully eclipse map WASP-43b \citep{hammond2024}, from which, combined with the phase information out of eclipse, tight constraints on the longitudinal hotspot offset were able to be derived. This shift of the hotspot from the substellar point is induced by thermal forcing from the extreme day-night temperature contrast, and is predicted by a number of hot Jupiter GCMs \citep{showman2011}. The magnitude of the shift is therefore a key diagnostic of the atmospheric dynamics of hot Jupiters. Eclipse mapping grants us new insight into these dynamics, allowing us to better our understanding of the multidimensional processes at play in these highly dynamic environments.

The structure of this paper is as follows. In Section \ref{sec:obs}, we present the observations, data reduction, and construction of the emission spectrum. In Section \ref{sec:modelling}, we present our forward model interpretation of the emission spectrum. In Section \ref{sec:mapping}, we detail our eclipse mapping methodology and results. Finally, we present our conclusions in Section \ref{sec:conclusions}.

This paper is part of a series by the JWST Telescope Scientist Team (JWST-TST)\footnote{\url{https://www.stsci.edu/~marel/jwsttelsciteam.html}} which uses Guaranteed Time Observer (GTO) time awarded by NASA in 2003 (PI Matt Mountain) for studies in three different subject areas: (a) Transiting Exoplanet Spectroscopy (lead: N. Lewis); (b) Exoplanet and Debris Disk High-contrast Imaging (lead: M. Perrin); and (c) Local Group Proper Motion Science (lead: R. van der Marel). Previously reported results across these areas include: (a) \citet{miritransit}, A. Gressier et al. (accepted, AJ), and D. R. Louie et al. (in review, AJ); (b) \citet{ruffio2023}, \citet{rebollido2024}, \citet{hoch2024}, and \citet{kammerer2024}; and (c)  \citet{libralato2023}. A common theme of these investigations is the desire to pursue and demonstrate science for the astronomical community at the limits of what is made possible by the exquisite optics and stability of JWST. The present paper is part of our work on Transiting Exoplanet Spectroscopy, which focuses on detailed exploration of three transiting exoplanets representative of key exoplanet classes: Hot Jupiters (WASP-17b, GTO~1353), Warm Neptunes (HAT-P-26b, GTO~1312), and Temperate Terrestrials (TRAPPIST-1e, GTO~1331). Here we present our observational analysis and interpretation of the MIRI/LRS 5--12 $\upmu$m dayside emission spectrum and eclipse map of WASP-17b.

\section{Observations, Data Reduction, and Spectrum Generation} \label{sec:obs}
We observed a single secondary eclipse of WASP-17b using JWST's Mid-InfraRed Instrument (MIRI) Low Resolution Spectrometer (LRS) in Slitless Mode \citep{kendrew2015} on UT 2023 March 14 (program observation 6). This instrument samples the 5--12 $\upmu$m range at an average native spectral resolution of $R\sim100$. The exposure time totalled 9.92 hr, covering the 4.42 hr eclipse \citep{anderson2011} with ample pre- and post-eclipse baseline. Due to the length of the observation, it was divided into two exposures, each comprising 638 integrations (1276 total), with each integration being 27.99 s in length and comprising 175 groups. The data were read out using the FASTR1 readout pattern, and they are provided by the JWST archive\footnote{\dataset[DOI:10.17909/cqsf-sx69]{ http://dx.doi.org/10.17909/cqsf-sx69}} in 12 data segments to manage file size.

For comparison and validation purposes, we performed independent reductions of the data using two separate pipelines: \texttt{ExoTiC-MIRI} \citep{grantexoticmiri} and \texttt{Eureka!} \citep{eureka}, both of which are open-source. The details of each independent reduction are outlined below.

\subsection{Data Reduction: \texttt{ExoTiC-MIRI}}
\texttt{ExoTiC-MIRI}\footnote{\url{https://exotic-miri.readthedocs.io/}} interoperates with the standard STScI \texttt{jwst} pipeline \citep{Bushouse2022}, optionally swapping in certain custom reduction steps that have been designed specifically for the reduction of MIRI/LRS time-series data. We use version 1.8.2 of the \texttt{jwst} pipeline in this work with Calibration Data Reference System (CRDS) context map 1063. A decision tree of reductions was crafted, testing the effects of employing different custom steps and altering step parameters, in order to determine the optimal reduction pathway. By inspection of the morphology and consistency of the derived stellar spectra across the time series, and minimisation of the standard deviation in measured flux, we determined that the methodology employed in the \texttt{ExoTiC-MIRI} reduction of the MIRI/LRS transit of WASP-17b also gave the best results for the eclipse presented here. We outline these steps below, and refer the interested reader to \citet{miritransit} for further details.

Starting from the Stage 0 \textit{\_uncal.fits} files, our modified Stage 1 employed all the standard \texttt{jwst} pipeline steps appropriate for MIRI/LRS time-series data except for the linearity correction step, and with the modifications of increasing the jump step rejection threshold to $15 \sigma$ to avoid spurious flagging of cosmic rays, and the gain value to 3.1 electrons/DN per \citet{bell2023first}. With regard to the linearity correction step, similar group-level features were present as those observed in \citet{miritransit}, namely: residual non-linearities across all groups, likely as a result of the brighter-fatter effect \citep[BFE;][]{argyriou2023brighter}; evidence of the reset switch charge decay (RSCD) effect impacting the first 12 groups of each integration; and  evidence of the last frame effect impacting the final group of each integration \citep{ressler2015mid, wright2023mid}. To account for the combined impact of these effects, we applied the same custom \texttt{ExoTiC-MIRI} linearity correction as \citet{miritransit}: the first 12 and the last group of each integration were removed altogether, and then a linear fit was extracted from the 12th-40th groups (which we determined to be the least impacted by the aforementioned systematics) and used to extrapolate a linearity correction factor to the remaining groups. The outputs of this stage are rate images in the format of \textit{\_rateints.fits} files.

In Stage 2, the default \texttt{jwst} versions of the assign world coordinates, source type, and flat-field steps were applied to the \textit{\_rateints.fits} files. From \texttt{ExoTiC-MIRI}, the following steps were run. The background was subtracted using the ``row-by-row'' method, which calculates the background in each row of the detector (i.e., for each wavelength) by fitting them with an $n$th order polynomial. We defined the background region in the 12th$-$22nd and the 50th$-$68th pixel columns, respectively, on either side of the spectral trace, and fit it using a fourth-order polynomial. We found that our chosen method of background subtraction had the most significant impact on the quality of the derived 1D stellar spectra, and in particular that performing a constant background subtraction led to negative fluxes in the light curves, as the background can vary significantly over this large wavelength range. A fourth-order polynomial suited to best match the complexity of the background without leading to over- or under-fitting, and performing the correction row-by-row was necessary in order to properly model the wavelength-dependence of the background. We next performed outlier cleaning by estimating a spatial profile from polynomial fits to the detector columns as per optimal extraction \citep{horne1986}, iteratively replacing $>$5$\sigma$ outliers from this profile along with pixels with the data quality flag \textsc{DO\_NOT\_USE}. This step was carried out both pre-background subtraction so that outliers did not significantly skew the background count, and post-background subtraction to catch any remaining outliers. Overall, we found the detector to be very well-behaved, with only 0.3\% of pixels identified as outliers by this algorithm. Finally, the time series of stellar spectra were extracted from the spectral trace via the ``box extraction'' method using a fixed-width top-hat aperture; this aperture was centered on the 36th column with a width of $\pm3$ pixels, the latter of which was determined via minimisation of the standard deviation of measured flux. The relative shifts of the spectral trace in x- and y-position were determined via cross-correlation with the median position across the time series, for later decorrelation in systematic fitting.

From this time series of stellar spectra, we constructed a broadband 5--12 $\upmu$m white light curve, along with spectroscopic light curves using two binning regimes with bin widths of 0.25 $\upmu$m and 0.50 $\upmu$m, respectively.
Due to the lower signal of eclipse observations in comparison to transit, 0.25 $\upmu$m was deemed to be the finest feasible bin size, compared to 0.125 $\upmu$m for \citet{miritransit}, and the purpose of the two binning regimes is to validate that our results are not biased by the ``odd-even row effect'' known to affect JWST MIRI observations \citep{ressler2015mid}.

\begin{figure*}
\includegraphics[width=0.99\textwidth]{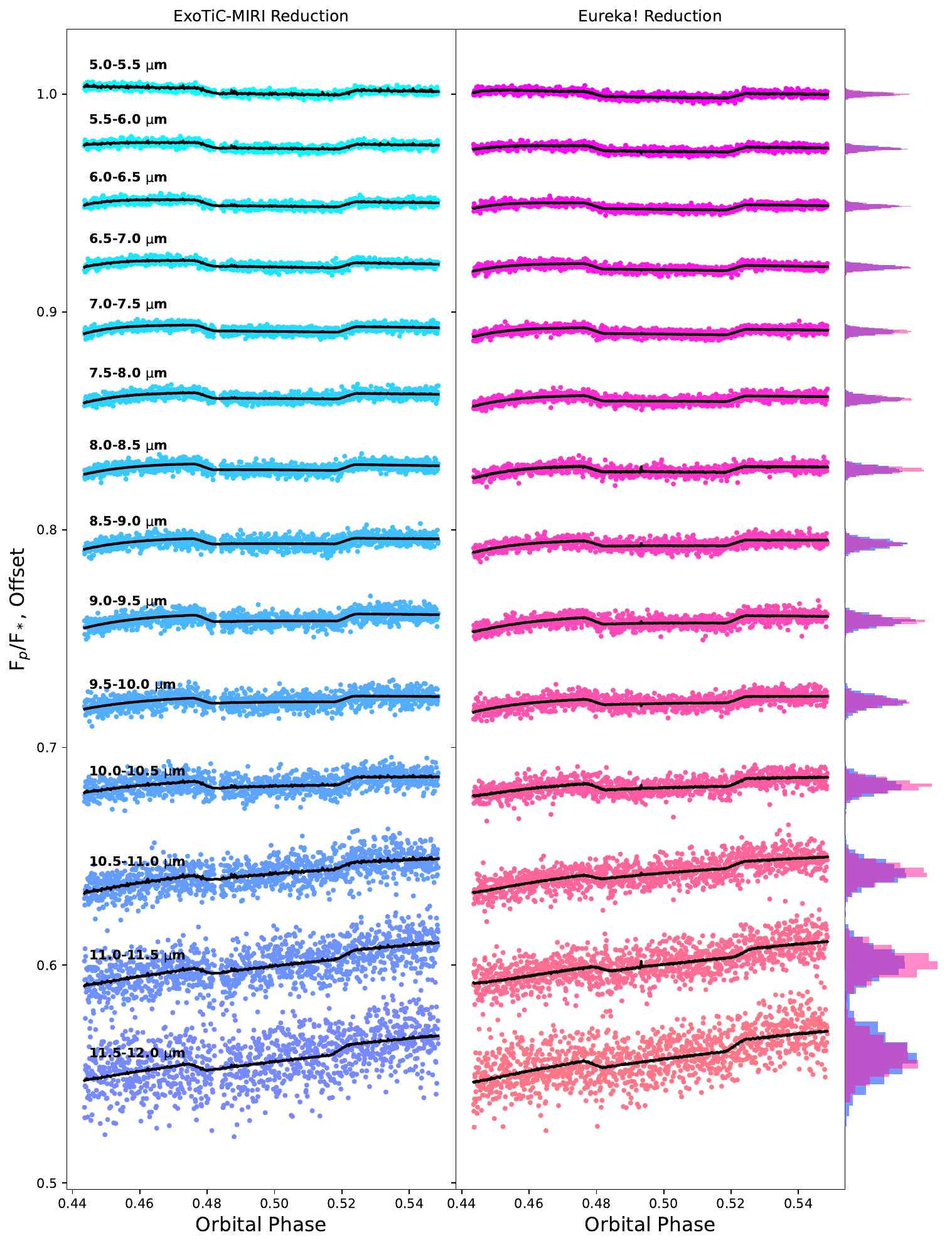}
\caption{\texttt{ExoTiC-MIRI} and \texttt{Eureka!} 0.50 $\upmu$m binned spectroscopic light curves, excluding any trimmed or masked integrations (see text for details), with best-fit models overlaid. The histograms along the right-hand y-axis show the residuals between the data and model fits.}
\label{fig:waterfallplot}
\end{figure*}

\subsection{Data Reduction: \texttt{Eureka!}}
We performed an additional reduction of our WASP-17b MIRI/LRS eclipse observations using the \texttt{Eureka!} (v0.9)\footnote{\url{https://eurekadocs.readthedocs.io/}} pipeline \citep{eureka}, also in a manner similar to that outlined using the same pipeline for the MIRI/LRS transit of WASP-17b in \citet{miritransit}. We employ \texttt{Eureka!}
stages 1--4 to extract our light curves starting from the {\it uncal.fits} files. 
Eureka! stages 1 and 2 leverage the JWST Science Calibration Pipeline \citep[v1.8.2][\texttt{jwst}]{Bushouse2022}. 
We largely use the default MIRI time-series observations settings for stages 1 and 2 in \texttt{Eureka!}, but find that a 15$\sigma$ jump 
rejection threshold is needed to balance cosmic-ray rejection and preserve our astrophysical signal.

In stage 3, we perform row-by-row background subtraction using a 10 pixel half-width exclusion region relative to the source position to determine the background, along with outlier cleaning using a double-iteration 5$\sigma$ rejection threshold (as calculated from the standard definition of standard deviation) from the median frame, which identified $<1$\% of pixels as being outliers. Finally, we carry out optimal spectral extraction on the generated calibrated images ({\it calints.fits}) using a 4 pixel 
half-width extraction aperture. 
The spatial profile position ($x_{\mathrm{pos}}$ for MIRI/LRS) and width ($x_{\mathrm{width}}$ for MIRI/LRS) as a function of time are determined 
from cross-correlation with the median frame. With our per wavelength/pixel row per time stellar spectra extracted in stage 3, we then bin the spectra into a prescribed number of spectroscopic channels with the same bin widths as employed in \texttt{ExoTiC-MIRI} and generate white light (5--12~$\upmu$m) and spectroscopic light curves. Further sigma-clipping is performed on the time series using a box-car filter with a width of 10 points and a 5$\sigma$ threshold. Our sigma-clipping removes 1.25\%, 0.13\%, and 0.22\% of the total data points for the white light curve, 0.25 $\upmu$m bin, and 0.50 $\upmu$m bin spectroscopic light curves, respectively. The higher value for the white light curve can be attributed to an anomalous rise in flux comparable to the eclipse depth present in $\sim$20 integrations immediately post-ingress. This feature, which is of as yet undetermined origin, was most evident in the white light curve due to the lower scatter compared to the spectroscopic light curves.

\subsection{Light-curve Fitting and Spectral Generation: \texttt{ExoTiC-MIRI}}
The light curves were fit using the model
\begin{equation}
    f(t) =  E(t_s, \theta) \times S(t_{s}, t_{m}, x, y),
\end{equation}
where $E$ is the astrophysical eclipse model with parameter vector $\theta$, generated using \texttt{BATMAN} \citep{kreidberg2015batman}, and $S$ is the coupled systematic model, given by

\begin{equation}
    S(t, x, y) = f_{0} + r_0 e^{-r_1 t_{s}} + c_{0} t_{m} + s_{0} |x-\bar{x}|y,
\label{eq:exotic_miri_light_curve_systematics}
\end{equation}

where $t_{s}$ is the observation times minus the start frame time (accounting for trimmed integrations at the start of the light curve), $t_{m}$ is the observation times minus the predicted centre of eclipse time (both in units of days), $f_{0}$ is a baseline offset, $|x-\bar{x}|$ is the absolute $x$ (spatial) shifts of the spectral trace on the detector relative to the median location on the detector, $y$ is the respective y (spectral) shifts, and $r_0$, $r_1$, $c_0$, and $s_0$ are coefficients to be fit.

The astrophysical model fit for eclipse depth, $F_P / F_*$, and time of central eclipse, $t_e$, in both the white light and spectroscopic light-curve fitting, whilst all other physical parameters were held fixed to the values in Table \ref{tab:system_and_fitting_params}. The predicted centre of eclipse time was set as 0.5 phase after the \texttt{ExoTiC-MIRI} centre of transit time determined by \citet{miritransit}, wherein the observed transit directly preceded the eclipse observed in this data. The time of central eclipse was not fixed to the value from the white light curve fit when fitting the spectroscopic light curves because we expect to see a wavelength-dependence to this value: different wavelengths probe different pressures in the atmosphere, and if the dynamics are different at these different pressures, then the hotspot location may be altered, and therefore so too will the timing of central eclipse in that channel \citep{eclipse_timing}. These variations with wavelength are thus a diagnostic of the dynamics of the atmosphere as a function of pressure.

We crafted a grid of systematic models, with varying inclusions of one or two exponentials, order of the polynomial in time, ways of treating shifts of the spectral trace, and testing the inclusion of detector mnemonics, and fit them to the white light curve using least-square minimisation (LSM) in order to determine the optimal model.
The choice of systematic model is important for MIRI/LRS light-curve fitting due to the strong systematics \citep{bell2023first}. The systematic model that was found to best minimise the residuals and red noise whilst maintaining minimal degeneracies between the systematic parameters is described in Equation \ref{eq:exotic_miri_light_curve_systematics}, and includes terms that describe the exponential ramp at the start of the observation (parameterised by $r_0$ and $r_1$), a linear evolution with time (parameterised by $c_0$), and response variations with trace position on the detector (parameterised by $s_0$). This is similar to other analyses of MIRI/LRS time-series observations of transiting exoplanets \citep[e.g.][]{bell2023first, miritransit}.
For the latter term, the absolute values of the $x$ shifts relative to the median across all integrations,  $|x-\bar{x}|$, were used so as not to bias whether a shift left or right on the detector directly corresponds to an increase or decrease in flux.

The first 65 integrations ($\sim$30 minutes) of each light curve, corresponding to the most severe portion of the ramp, were excluded from the fit in order to ease the systematic fitting. The first integration of each of the 12 data segments were also masked due to anomalous background levels, which was similarly observed in our \texttt{Eureka!} reduction and by \citet{miritransit}.
We also observe the same anomalous rise in flux in the $\sim$20 integrations post-ingress as the \texttt{Eureka!} reduction. The fitting was carried out both with and without these 20 integrations masked, and whilst the fit itself was largely invariant across the channels, calculation of the Allan deviation \citep{allan_dev} showed that excess correlated noise remained in the white light curve when including these integrations, and so they were also masked in the final fit.

\begin{figure*}
\includegraphics[width=0.99\textwidth]{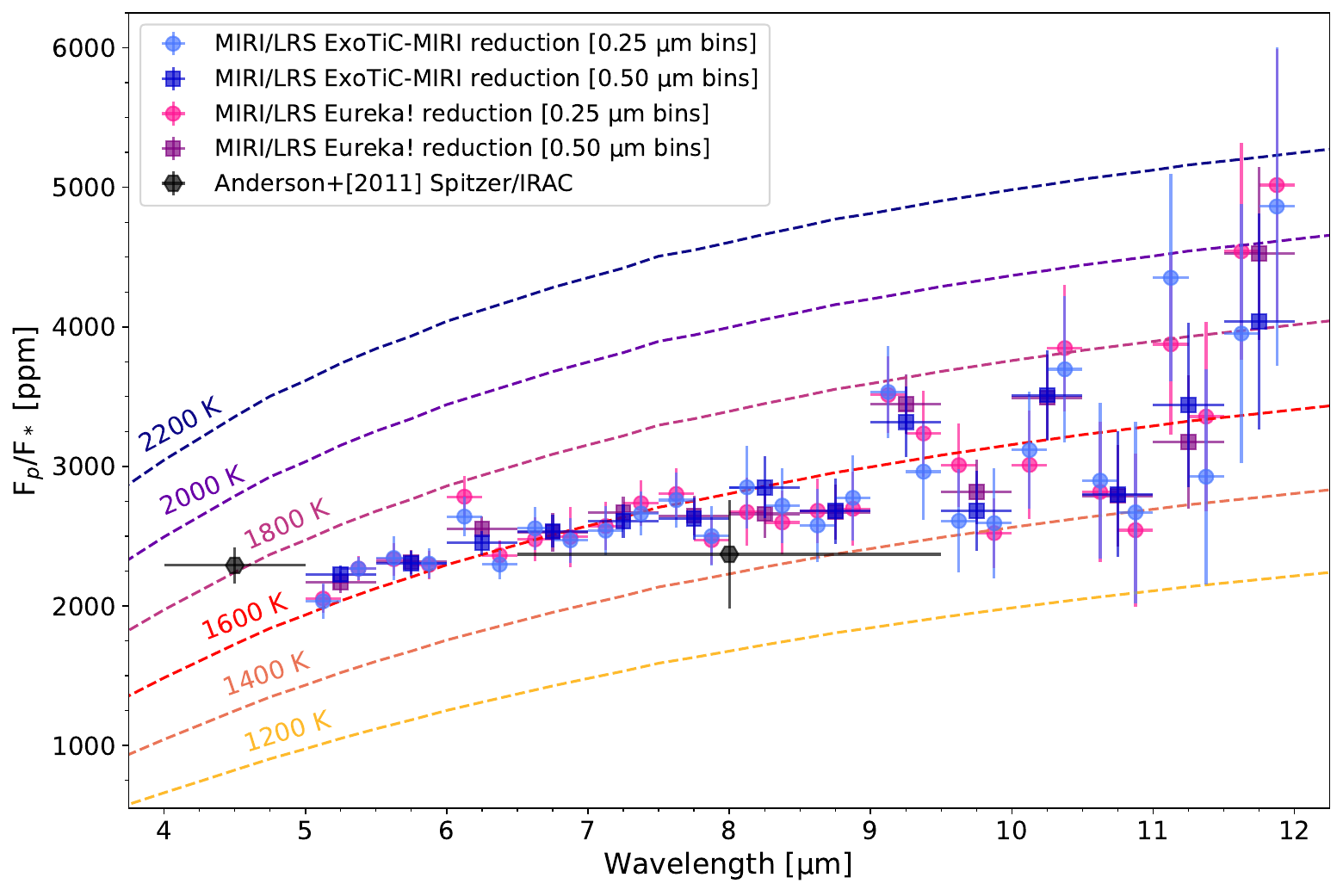}
\caption{The 5--12 $\upmu$m emission spectrum of WASP-17b generated using \texttt{ExoTiC-MIRI} (0.25 $\upmu$m bins in light blue circles, 0.50 $\upmu$m bins in dark blue squares) and \texttt{Eureka!} (0.25 $\upmu$m bins in pink circles, 0.50 $\upmu$m bins in purple squares). The black hexagons are previous Spitzer measurements \citep{anderson2011}. The dashed lines are blackbodies corresponding to temperatures ranging from 1200 to 2200 K, to give a scale of the temperatures probed by this data and to highlight deviations due to spectral features. All data products and models are available at \url{https://doi.org/10.5281/zenodo.12571830}.}
\label{fig:spectra}
\end{figure*}

Because of the increasingly large scatter at longer wavelengths, LSM resulted in overly large uncertainties on the fit parameters for these channels, from which it would be difficult to derive meaningful constraints. We interpret this as being due to the LSM routine failing to properly converge on the global probability minima for these lower signal-to-noise (SNR) channels. To alleviate this, the Markov Chain Monte Carlo (MCMC) routine \texttt{emcee} \citep{foreman2013emcee} was implemented in order to better explore the parameter space, and so that priors could be used to guide the fit (see Table \ref{tab:system_and_fitting_params} for values). The optimal parameters determined from the LSM fit to the white light curve were used to initialise the MCMC because it has the highest SNR, with wide priors to confirm that, if the same solution was found from the latter, then the former was in fact converging on the global solution and not getting trapped in a local probability minima. This was found to be the case, and so these same values were used to initialise the MCMC sampling of the spectroscopic light curves with the same wide priors. The prior on the centre of eclipse time was set to be approximately half the length of ingress/egress, $\sim$15 minutes, as this is the maximum amount one would expect that a hotspot offset could feasibly shift the eclipse timing in either direction. We also compared the LSM and MCMC fits to the spectroscopic light curves in order to ensure that the MCMC priors do not bias our solutions at lower SNR. We found the routines to agree very well within their uncertainties, indicating that no bias is present, with the primary difference being that the MCMC accomplishes our aim of better constraining the fit parameters due to its better convergence on the global probability minima, which is testified by its higher model evidence compared to the LSM fits. See Table \ref{tab:system_and_fitting_params} for a comprehensive summary of system parameters and prior distributions used in both the \texttt{ExoTiC-MIRI} and \texttt{Eureka!} light-curve fitting.

Due to its faster runtime, LSM was still employed prior to the MCMC sampling in order to perform outlier cleaning. Integrations $>$4$\sigma$ discrepant from the model fit were identified as outliers, iteratively removed, and the sampling re-performed until none remained. This resulted in at most one or two data points being removed in only a handful of cases, totalling 0.01\% and 0.03\% for the 0.25 $\upmu$m and 0.50 $\upmu$m spectroscopic light curves, respectively, and none for the white light curve. These lower percentages compared to the \texttt{Eureka!} values can be attributed to our manual exclusion of the anomalous data points at the start of each data segment and post-ingress prior to outlier trimming, whereas \texttt{Eureka!} does this procedure in Stage 4, prior to light-curve fitting, wiht no a priori manual exclusion.

An error multiplier, $\beta$, was additionally included in the MCMC fit to account for possible error misestimation, in particular as a result of detector gain uncertainty \citep{bell2023first}. The MCMC was run with 64 walkers for 10,000 steps, the first 2500 of which were discarded as burn-in. We employed the Gelman-Rubin statistic, $\hat{r}$, to test for converge, with the criterion of $\hat{r}<1.1$, as is appropriate for lower-SNR data \citep{gelman_rubin}.

\subsection{Light-curve Fitting and Spectral Generation: \texttt{Eureka!}}
\texttt{Eureka!} uses the same astrophysical model as \texttt{Exotic-MIRI}. 
The systematics model employed in our \texttt{Eureka!} light-curve fitting is similar to \texttt{ExoTiC-MIRI} except for the way that we model shifts of the spectral trace on the detector. Our systematic model is of the form

\begin{multline}
S(t_s, t_m, x_{pos}, x_{width}) =f_0+c_0 t_m +r_0 e^{-r_1t_s} \\ +x_0x_{pos} +x_1x_{width}, \label{eureka_sys}
\end{multline}

where $f_0$, $c_0$, $r_0$, and $r_1$ are the same coefficients to be fit as in \texttt{ExoTiC-MIRI}, with similarly defined time arrays, but $x_0$ and $x_1$ are our different fit coefficients for detrending against spectral trace shifts.
The term $x_{\mathrm{pos}}$ represents the spatial drift, and the term $x_{\mathrm{width}}$ represents changes in the PSF width. We similarly trim the first 65 integrations ($\sim$30 minutes) of each light curve.
Fit alongside these systematic coefficients and the astrophysical model is an error inflation term ($\beta$). See Table \ref{tab:system_and_fitting_params} for a summary of the employed priors. We fit the \texttt{Eureka!} light curves using the MCMC sampling algorithm \texttt{emcee} \citep{foreman2013emcee} with 3000 steps, 200 walkers, and 1500 steps for burn-in. We employed the same convergence metric, the Gelman-Rubin statistic, as \texttt{ExoTiC-MIRI}, with the same criterion of $\hat{r}<1.1$ \citep{gelman_rubin}.

\subsection{\texttt{ExoTiC-MIRI} and \texttt{Eureka!} Intercomparison}
\label{sec:reduction_comp}
The 0.50 $\upmu$m bin spectroscopic light curves from each reduction are shown in Figure \ref{fig:waterfallplot} with the best-fit models overlaid. The evolution of the systematics across the channels are particularly evident: the linear slope increases in severity going to longer wavelengths, and the ramp transitions from a decaying morphology at the shortest wavelengths to an increasingly rising morphology beyond 5.5 $\upmu$m. This transition occurs at much shorter wavelengths than has been observed for other MIRI/LRS observations \citep[e.g.][]{bell2023first, zhang2024}, wherein the transition occurs in the ``shadowed region" of the detector beyond 10.5 $\mu$m. This highlights the variable nature of the ramp between observations, the exact mechanisms of which are still not well understood. The histograms along the right-hand y-axis show the residuals between the data and model fits, highlighting the increasing scatter to long wavelengths. Overall, we find the light curves to be well fit and the systematics to be correctable across the channels, with no evident residual correlated noise in the Allan deviation plots (see online supplementary material).

The 0.25 and 0.50 $\upmu$m resolution emission spectra derived from each reduction are shown in Figure \ref{fig:spectra}. The black data points are previous Spitzer/IRAC measurements \citep{anderson2011}, the values of which are broadly consistent with those that we derive from our MIRI/LRS measurements.
The coloured dashed lines are blackbody curves of varying temperatures, to give a scale of the temperatures probed by this data and to highlight deviations due to spectral features. These were generated from blackbodies computed using the Planck function at the specified temperature, and converted to eclipse depth units by dividing through by the stellar spectrum and multiplying by the ratio of the planetary-to-stellar radii squared, $(R_{p}/R_{*})^2$. We generated the stellar spectrum by interpolating in a grid of pre-calculated Phoenix spectra \citep{phoenixmodels} with WASP-17A parameters from \citet{southworth2012wasp17a}, namely
$\textup{T}_{{\textup{eff}}}$=6500$\pm$75 K, $[\textup{Fe/H}]$=-0.25$\pm$0.09, and $\textup{log(g)}$=4.149$\pm$0.014. For the value of $(R_{p}/R_{*})^{2}$, we adopt the \texttt{ExoTiC-MIRI} value for the blackbody curves displayed in Figure \ref{fig:spectra}, but for determination of the best-fit blackbody temperature of each independently reduced spectrum (i.e., \texttt{ExoTiC-MIRI} vs. \texttt{Eureka!}), we adopt the corresponding $(R_{p}/R_{*})^{2}$ as used in the respective light-curve fitting (see Table \ref{tab:system_and_fitting_params}).

These blackbody fits give us highly consistent best-fit blackbody temperatures between the different reductions of $1604 \pm 16$ K for the \texttt{ExoTiC-MIRI} spectrum and $1607 \pm 18$ K for the \texttt{Eureka!} spectrum, both at 0.25 $\upmu$m resolution. When additionally accounting for uncertainties in the stellar parameters, these blackbody uncertainties increase to give values of $1604^{+25}_{-31}$ K for \texttt{ExoTiC-MIRI} and $1607^{+27}_{-34}$ K for \texttt{Eureka!}. We do not account for uncertainties induced by our choice of stellar model framework (i.e., using a Phoenix model rather than another stellar model architecture), as this is an inherent part of data interpretation. We elect to use Phoenix models as this is the standard in the field, but we note that most well-vetted stellar models have been benchmarked against one another and will give consistent results within uncertainties in the data itself. These temperatures are $\sim$200 K cooler than those determined from the NIRISS/SOSS eclipse of WASP-17b presented by A. Gressier et al. (accepted, AJ), which is to be expected since those data probe deeper pressures in the atmosphere. The relatively small uncertainties on our derived blackbody temperatures are due to the fact that the shorter wavelengths ($<$8 $\upmu$m), which have smaller eclipse depth uncertainties and therefore carry the greatest weight in the fit, largely follow a blackbody curve. There is very little sway for the blackbody fit to adjust to similarly fit the longer-wavelength deviations whilst still adequately fitting the shorter-wavelength channels, because its shape is fixed; the only freedom of the model is to shift up and down in eclipse depth. Nevertheless, these temperature values are not to be taken with too much weight because they only give an approximate idea of the average atmospheric temperatures probed; in reality, we see deviations from blackbody curves due to spectral features in the atmosphere of the planet (see Figure \ref{fig:spectra}).

The \texttt{ExoTiC-MIRI} and \texttt{Eureka!} spectra are better than 1$\sigma$ consistent throughout, in particular at the shorter wavelengths ($<$8 $\upmu$m). The median residuals between the 0.25 $\upmu$m resolution spectra (10 ppm) is $<$0.05$\sigma$ that of the median uncertainty in the eclipse depths (284 ppm for \texttt{ExoTiC-MIRI}, 236 ppm for \texttt{Eureka!}) across the wavelength range. The higher median eclipse depth uncertainties for the \texttt{ExoTiC-MIRI} spectrum can be largely attributed to the data points $>$10 $\upmu m$ (this can be seen by the wider extent of the histograms in Figure \ref{fig:waterfallplot} at these wavelengths); removing these gives more similar median uncertainties of 188 ppm for \texttt{ExoTiC-MIRI} and 182 ppm for \texttt{Eureka!}. The marginally higher precision of the \texttt{Eureka!} reduction is to be expected because \texttt{ExoTiC-MIRI}, in using a smaller subset of the groups to perform ramp fitting in Stage 1 of the data reduction, sacrifices a degree of precision, with the aim of increasing accuracy \citep{miritransit}. \texttt{Eureka!} also employs optimal extraction, which was found to give higher precision than a box aperture, as employed in \texttt{ExoTiC-MIRI}. These factors are reflected in the median precision of 445 ppm achieved in the \texttt{ExoTiC-MIRI} white-light-curve fit, compared to 421 ppm for the \texttt{Eureka!} fit. Both of these compare well to photon noise, however, at 1.08$\times$ and 1.02$\times$, respectively.

\begin{figure*}[hbt!]
\includegraphics[width=0.99\textwidth]{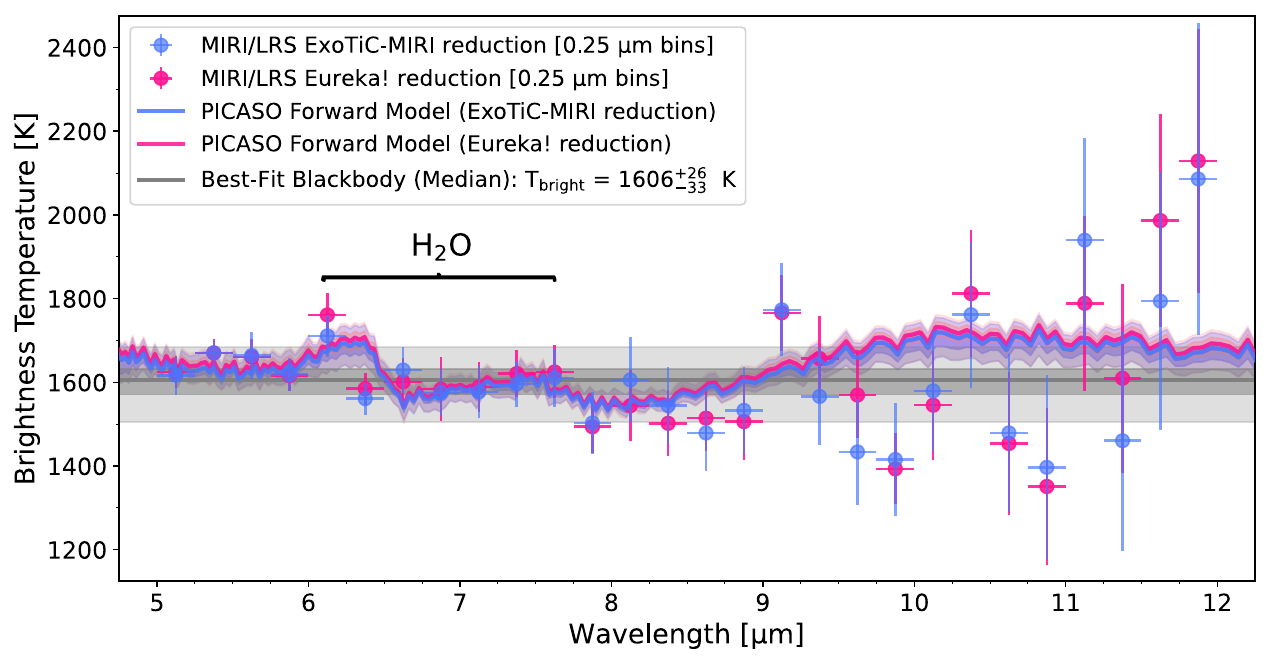}
\caption{PICACO forward-model fits to the 0.25 $\upmu$m resolution brightness temperature spectra from both reduction pipelines (\texttt{ExoTiC-MIRI} in blue, \texttt{Eureka!} in pink). The shaded regions are the 1$\sigma$ and 3$\sigma$ credibility regions. The free parameters in the forward models fits are interior temperature ($T_{\mathrm{int}}$), heat redistribution efficiency ($f_{\mathrm{rd}}$), metallicity (log $\mathrm{[M/H]}$ $\times$ Solar), and carbon-to-oxygen ratio ($\mathrm{C/O}$ ($\times$ Solar =0.458)). We highlight the wavelengths where we observe absorption due to water. We additionally show the median best-fit blackbody to both spectra in grey with shaded 1$\sigma$ and 3$\sigma$ credibility regions.}
\label{fig:forward_model}
\end{figure*}

We note that slightly different system parameters were adopted for light-curve fitting for the separate reductions, these being the ratio of the semi-major axis to stellar radius, $a/R_{*}$, inclination, $i$, and the ratio of the planetary to stellar radius, $R_{p}/R_{*}$ (see Table \ref{tab:system_and_fitting_params}). Both reductions employ the same values as used in the fits to the same respective reductions of the MIRI/LRS transit of WASP-17b presented by \citet{miritransit}. The consistency of the spectra, despite these minor differences, further demonstrates the robustness of our respective reduction methodologies. Additionally, the consistency between both binning regimes, even at wavelengths where the 0.25 $\upmu$m bin includes an odd number of pixels and hence the 0.50 $\upmu$m bin includes an even number, lends further credence to the reliability of our results and validates that they are not significantly impacted by the ``odd-even row effect'' \citep{ressler2015mid}.

\section{Atmospheric Forward Modelling} \label{sec:modelling}

\subsection{Method}\label{sec:modelling_method}
In order to compute atmospheric forward models, we directly follow the methodology of \citet{grantexoticmiri} for the PICASO radiative-convective thermochemical equilibrium modeling. Moreover, we use the initial WASP-17b grid computed from \citet{grantexoticmiri} but include additional models to account for WASP-17b’s hotter dayside, as opposed to the limb-averaged spectra presented there. For completeness, we reiterate the initial methodology here, and provide additional context for how we expanded the grid.

PICASO is an open-source atmospheric forward and inverse model \citep{batalha_picaso, mukherjee2023picaso}. As opposed to \citet{grantexoticmiri}, which used PICASO v3.1, we used PICASO v3.2\footnote{\url{https://github.com/natashabatalha/picaso}}, which included specific updates that allowed us to run models for higher interior temperatures, as is indicated by our program's NIRISS/SOSS emission spectrum of WASP-17b (A. Gressier et al. accepted, AJ). This planet-specific climate grid was computed without clouds for a total of four parameters: interior temperature (200, 300, \textbf{400, 600, 800} K, \citealt{thorngren2019intrinsic, sarkis2021evidence}), atmospheric metallicity (9 values between 1--100$\times$Solar), C/O ratio (5 values between 0.25--2$\times$Solar, where Solar=0.458), and the heat redistribution factor (0.5, 0.6, 0.7, 0.8, \textbf{0.9, 0.95}, where 0.5=full redistribution and 1.0=no redistribution). Bolded values are new grid points added for this specific analysis. Briefly, the PICASO climate solver is based on the methodology developed in \citet{marley1999thermal}, and the correlated-K opacities detailed in \citet{Marley2021} and released on Zenodo \citep{lupu_roxana_2021_5590989}. To compute high resolution spectra, we computed emission spectra for the entire grid (since \citet{grantexoticmiri} focused on transmission geometry). We used the \citet{Toon1989} source function technique with opacities resampled at $R=15,000$ \citep{natasha_batalha_2020_6928501} down from their original resolution ($R\sim10^{6}$) \citep{freedman2008opacities, nezhad21}. The dominant source of opacity in the MIRI/LRS bandpass is H$_2$O, for which we used the line list derived in \citet{Polyansky2018H2O}. We use a fitting scheme identical to that outlined in \citet{grantexoticmiri}, which leverages the ``MLFriends'' nested sampling Algorithm \citep{MLFriends2016, MLFriends2019}, which is embedded in the open-source Ultranest code \citep{Ultranest}. 

We conduct the fit to the data in brightness temperature rather than eclipse depth because in this basis, deviations from a flat line (i.e., a fixed temperature) highlight spectral features in the atmosphere. This matches the architecture of transmission spectroscopy, which is more compatible with the modelling framework. We convert the emission spectra MIRI/LRS data from units of eclipse depth to brightness temperature using

\begin{equation}
    T_{b} = \frac{hc}{\lambda k} \mathrm{log}\left(1+ \frac{2 \pi hc^{2} (R_{P}/R_{*})^{2}} {\lambda^{5} F_{*} (F_{p}/F_{*})} \right),
\end{equation}

\noindent where $\lambda$ is the central wavelength of the channel, and the stellar flux, $F_{*}$, is modelled using the same Phoenix model as used to generate the blackbody functions in Figure \ref{fig:spectra}. The factor of $\pi$ in the numerator is to account for the fact that the Phoenix model values are calculated per steradian. We use the respective \texttt{ExoTiC-MIRI} and \texttt{Eureka!} values for $R_{p}/R_{*}$ as derived in \citet{miritransit}; see Table 1. We bootstrap the eclipse depth uncertainties to brightness temperature uncertainties from 10,000 random draws of the posterior distribution.

\subsection{Results}\label{sec:modelling_results}
Figure \ref{fig:forward_model} shows the median best-fit PICASO forward models to both the \texttt{ExoTiC-MIRI} and \texttt{Eureka!} 0.25 $\upmu$m resolution brightness temperature spectra. As is evident from the overlap between the models, highly consistent results are found between the reductions. We also find consistent results between the different resolution spectra (0.25 $\upmu$m and 0.50 $\upmu$m, the latter of which can be found in the online supplementary material), further demonstrating that our results are not noticeably impacted by the ``odd-even row effect" \citep{ressler2015mid}. 

Of the four parameters that we differ in the model grid, we find that this MIRI/LRS emission spectrum is unable to adequately constrain the internal temperature, metallicity, or C/O ratio of WASP-17b. The latter two are not particularly surprising, given the limited chemical content of the MIRI/LRS bandpass in emission for high-temperature hot Jupiters. The former is also to be expected when considering the temperatures probed by this dataset, which indicate that we are probing lower pressures in the atmosphere where stellar irradiation effects will dominate over the internal heat of the planet. Because our posterior distributions are non-Gaussian for these parameters, we cannot derive meaningful uncertainties, but for completeness, we present the median values for each reduction along with the range of the distributions below.

We find consistent median internal temperatures between the reductions of 430 K, with ranges between 270 and 560 K. Given that we only explore values between 200 and 800 K, this range reflects our flat posteriors, which push up against the boundaries of the explored values. Despite our weak constraints, we note that these values are broadly consistent with the elevated internal temperature proposed by A. Gressier et al. (accepted, AJ) to at least partly explain the excess emission observed below 1 $\upmu$m in the NIRISS/SOSS emission spectrum of WASP-17b.

The metallicity is better constrained, with a peak in the posterior at a slightly supersolar value of $\sim$2$\times$Solar. However, this peak is fairly broad, and we additionally observe a tail end in the posterior distribution to much higher supersolar values, resulting in an overall range of plausible values between 1--30$\times$Solar. Again, whilst our constraints are weaker, these values are in general agreement with those observed in the MIRI/LRS transmission spectrum \citep{miritransit} and NIRISS/SOSS emission spectrum (A. Gressier et al. accepted, AJ) of WASP-17b.

For the C/O ratio, our posterior pushes up against the upper edge of the parameter space, resulting in tightly constrained values between 1.2 and 1.5$\times$Solar. However, we note that no carbon-bearing species are detected at significance in our spectra from either reduction; therefore, we cannot strictly draw meaningful conclusions from these values. Additionally, the non-Gaussian profile of the posterior distributions reveals the erroneously constrained nature of these values, further indicating their unreliability.

Whilst MIRI/LRS is not particularly rich in chemical information content at these high dayside temperatures, it is rich in thermodynamical content. We find precise heat redistribution factors between the \texttt{ExoTiC-MIRI} and \texttt{Eureka!} reductions of $0.93^{+0.01}_{-0.03}$ and $0.91^{+0.02}_{-0.01}$ (median $0.92\pm0.02$), with well-constrained Gaussian posteriors. These values are indicative of inefficient heat redistribution from the dayside to the nightside, which would produce a substantial day-night temperature contrast. This is in contrast to the highly efficient heat redistribution factor of 0.5 derived from the NIRISS/SOSS emission spectrum of WASP-17b (A. Gressier et al. accepted, AJ). However, the higher brightness temperatures derived from those results indicate that they are probing deeper in the atmosphere, where the \citet{kataria2016} GCM of WASP-17b shows a homogenised day-night thermal profile, compared to the large day-night contrast it shows at the temperatures probed by this dataset. Hence, these results are in fact consistent with one another, and in line with theoretical expectations. \citet{anderson2011} conversely infer an efficient heat redistribution in a similar bandpass to our MIRI/LRS results, but that result is only inferred from two photometric points, as opposed to our results which are derived from forward-model fitting to a comprehensive emission spectrum spanning a wider bandpass. \texttt{PICASO} fits to the MIRI/LRS transmission spectrum of WASP-17b also derive an inefficient heat redistribution factor, consistent with our emission spectrum results \citep{miritransit}.

\subsection{Unexplained Spectral Features}\label{sec:goodness_and_mystery_abs}
With regard to the goodness of these fits, the \texttt{ExoTiC-MIRI} model achieves a reduced chi-squared of $\chi_{\nu}$=1.22, in comparison to $\chi_{\nu}$=1.62 for the \texttt{Eureka!} model. Calculating the cumulative distribution function (CDF) for these model fits \citep{chi_squared_cdf}, we find that values of 0.73$<$$\chi_{\nu}$$<$1.27 fall within the 1$\sigma$ credibility region of the models, placing the \texttt{ExoTiC-MIRI} model just within this threshold and the \texttt{Eureka!} model far outside of it. We attribute this to the custom linearity correction employed in \texttt{ExoTiC-MIRI}, which aims to increase the accuracy of the results by extrapolating the correction from only the ``best-behaved" groups, but at the cost of a sacrifice in precision due to fewer number of groups employed in the eventual ramp-fitting, with the smaller uncertainties on the \texttt{Eureka!} spectrum placing the data points further outside the credible regions of the model.

We note that our best-fit blackbodies achieve comparable reduced chi-squareds of $\chi_{\nu}$=1.11 for \texttt{ExoTiC-MIRI} and $\chi_{\nu}$=1.64 for \texttt{Eureka!}. This is primarily because of the limited chemical information in the MIRI/LRS bandpass for such a high-temperature hot Jupiter. As previously noted, the gaseous species that is the dominant source of opacity in this bandpass is water, the absorption feature of which we observe and model at $\sim$6.5--7.5 $\upmu$m. However, the small magnitude of the feature is evident, hence the consistency of our dataset with a featureless atmosphere within the uncertainties of the data.

We do, however, highlight the regions of the spectrum beyond 9 $\upmu$m where we observe the potential presence of unexplained spectral features. It is in this region that the fit quality of the models decreases. Below 9 $\upmu$m, the \texttt{ExoTiC-MIRI} and \texttt{Eureka!} models achieve excellent reduced chi-squared values of $\chi_{\nu}$=1.04 and $\chi_{\nu}$=1.02, respectively. The inability of the models to fit the $>$9 $\upmu$m data adequately is what drives their poorer fit metrics overall. These wavelengths overlap with the ``shadowed region" that has been known to affect certain MIRI/LRS time-series observations \citep{bell2023first}, wherein residual correlated noise in these channels associated with the systematic ramp led to the necessary discarding of these long-wavelength data. We do not find this to be necessary for this dataset, with the ramp being correctable across the wavelength range and no residual correlated noise in any of the channels (see the online supplementary material for Allan deviation plots). We do not find these potential spectral features to be correlated with any uncorrected systematic, source of noise, or even detector mnemonic. In fact, the number of data points driving them and the consistency between the sloping features across the channels leads us to believe that these features may indeed be astrophysical.

Many plausible cloud species have absorption features at these wavelengths (see Figure \ref{fig:clouds}), namely silicate minerals including magnesium- and iron-based olivines. Silicate-based (quartz, $\mathrm{SiO_{2[s]}}$) clouds have already been detected on the limbs of WASP-17b using MIRI/LRS \citep{miritransit}, but the temperatures are likely too high on the dayside to sustain them in their condensate phase in large enough quantities to explain these features alone \citep[e.g.,][]{marley2013, helling2019}. Magnesium- and iron-based species, on the other hand, such as $\mathrm{Mg_{2}SiO_{4}}$ and $\mathrm{Fe_{2}SiO_{4}}$, can exist stably in the condensate phase at these temperatures \citep[e.g.,][]{visscher2010}; therefore, their presence or a mixture of these cloud species may explain these features, with the $\mathrm{Si}$ from the vaporised quartz clouds potentially going into their formation.

In order to quantify the significance of these long-wavelength features, we carry out a Gaussian significance test on them, as similarly employed in a number of JWST spectral studies \citep[e.g.,][]{Alderson2023_ERS,kirk2024}. The results of this analysis are presented in Figure \ref{fig:gauss_sig_test}. We subtract the PICASO forward model from the respective spectrum, essentially removing the ``known physics'' that we already model; deviations from a 0 K flat line therefore represent unmodelled features. In this basis, the well-modelled $<$9 $\upmu$m region of the spectrum is consistent with a 0 K flat line, but at long wavelengths, the deviation of the spectrum from the modelled physics is evident, with magnitudes of $\sim$400 K. Two Gaussian-like features are particularly evident: one at $\sim$9.8 $\upmu$m and another at $\sim$11 $\upmu$m (for the latter, we note that this could in fact be two features, but since this would only be driven by a single data point, we elect to model it as a single feature). Using the nested sampling package \texttt{dynesty} \footnote{\url{https://github.com/joshspeagle/dynesty}} \citep{dynesty}, we first fit these residual spectra as flat lines (shown in orange), with a prior centred on 0 K and a standard deviation, $\sigma$, of 100 K. We then re-fit the residual spectra with the inclusion of two Gaussians, with priors centred on the aforementioned approximated central wavelengths of these features with $\sigma=0.25$ $\upmu$m. The central value and $\sigma$ on the priors of the amplitude and standard deviation of these Gaussians are (-400, 250) K and (0.25, 0.25) $\upmu$m, respectively. The difference in model evidence, $\Delta$log(z), between these two fits therefore represents the statistical significance of these features. We find model evidences of $\Delta$log($z$)$=$2.67 for \texttt{ExoTiC-MIRI} and $\Delta$log($z$)$=$1.94 for \texttt{Eureka!}, which correspond to detection significances of 2.8$\sigma$ and 2.5$\sigma$, respectively, with the former's higher detection significance potentially being a result of the self-calibrated linearity correction. These values give strong indications of a likely astrophysical origin to these features, and therefore motivate further investigation.

In order to test whether or not clouds were able to fit these spectral features, we conducted a pseudo-retrieval exercise that utilized the cloud-fitting scheme developed in \citet{schlawin2024} and \citet{inglis2024}. Briefly, in addition to the four aforementioned cloud-free model grid parameters, we use the open-source Python code, \texttt{Virga} \footnote{\url{https://github.com/natashabatalha/virga}} \citep{batalha_virga, Rooney2022} to fit for: (1) cloud base pressure, (2) cloud sedimentation efficiency, which sets the altitude-dependent drop-off rate, (3) number density, which sets the maximum cloud optical depth, (4) effective cloud particle radius, and (5) the width of the cloud particle radius log-normal distribution. This methodology was shown to fit cloud features in the MIRI/LRS thermal emission of HD 189733 b (Inglis et al. accepted). However, when including these parameters in the forward-model fits to our spectrum, we were unable to adequately fit the features. This is because the best-fit cloud profile pushed the cloud base pressure toward higher pressures ($P_\mathrm{cld}>$1.5 bar). This means that the data itself did not adequately motivate any spectral contribution from the cloud.

We note, however, that due to the limited chemical content of the data, this cloud-fitting analysis was not comprehensive, using a generic prescription for clouds rather than modelling individual species. We also did not carry out a full retrieval analysis, as was required to identify the cloud species in \citet{miritransit}. Further work is required in order to determine the origin and potential physicality of these features, and this limited cloud analysis in no way excludes the possibility of that being the case. Our reasoning for not carrying out an exhaustive analysis is that, in order to properly identify the origin of these features, the inclusion of additional data, in particular at shorter wavelengths where we may observe additional signatures if these are in fact cloud species (e.g., their scattering properties in the optical), will be vitally helpful \citep{taylor2021}. This was the case in order to identify the $\mathrm{SiO_{2[s]}}$ clouds in the MIRI/LRS transmission spectrum, wherein the addition of optical HST data was crucial in order to break degeneracies between the cloud species and particle size \citep{miritransit}. Our program has constructed emission spectra for WASP-17b that span 0.6$-$12 $\upmu$m, which will be presented in H.R. Wakeford et al. (in preparation); these shorter-wavelengths datasets have a much higher chemical information content than MIRI/LRS for high-temperature hot Jupiters in emission, and so their inclusion with this dataset will better place us to identify the origin of these features, whether that be astrophysical or systematic noise. Therefore, we elect to devote the remainder of this analysis to that future work. For the remainder of this work, we instead now focus on the thermodynamical profile of WASP-17b, in which MIRI/LRS is most rich in information content.

MIRI/LRS is quickly emerging as a workhorse instrument in measuring the thermal emission profiles of hot Jupiters. In the next section, we outline our eclipse mapping efforts of the MIRI/LRS white light curve of WASP-17b. We use the \texttt{ExoTiC-MIRI} reduction as we have deemed it to give greater accuracy in the results due to the implementation of the self-calibrated linearity correction, as advocated by \citet{miritransit}.

\section{Eclipse Mapping} 
\label{sec:mapping}
We expect WASP-17b to have an appreciable eclipse mapping signal owing to its high temperature and comparably long ingress/egress duration of $\sim$30 minutes \citep{boone2024}. We use \texttt{ThERESA}\footnote{\url{https://theresa.readthedocs.io/}} \citep{theresa} to identify and extract such signals, focusing on the white light curve for maximum SNR, and transform them into a broadband eclipse map.

\subsection{Method} \label{sec: theresa_method}
Here, we briefly summarise the methodology of \texttt{ThERESA}, and refer the reader to \citet{theresa} for a more comprehensive explanation. In the traditional approach to eclipse mapping, one uses the complete and orthonormal set of spherical harmonics as basis maps because a linear sum of such bases can reproduce any brightness pattern on a sphere. To produce an eclipse map, the basis spherical harmonic maps are transformed into basis light curves and fit to the data as a linear sum. The weighting of each basis light curve gives the corresponding basis map weighting, and the overall eclipse map is the same linear sum of the basis maps. This method has had successful applications \citep[e.g,][]{majeau2012, dewit2012}, but is susceptible to signal degeneracy complications because, whilst the basis set of maps are orthogonal, the basis light curve components are not. Prior to JWST, this degeneracy problem was of least concern because maps could only be produced from a limited number of basis components, which are largely independent. However, since JWST light curves are far more precise and of high cadence, higher-order signals can now be identified and extracted, requiring a larger sum of basis-light curve fits, making them fraught with degeneracies, which can complicate interpretation.

\texttt{ThERESA} aims to overcome this signal degeneracy problem by transforming the basis maps and light curves to a new basis wherein the latter are also orthogonal. This is achieved by taking the basis set of spherical harmonic light curves and orthogonalising them using principal component analysis (PCA). The maps in this basis are referred to as ``eigenmaps'', and the light curves as ``eigencurves'' \citep{rauscher2018}. The same methodology that is used for spherical harmonic fitting is then applied to generate the eclipse map. \texttt{ThERESA} optimises for spherical harmonic degree, $l_{max}$ and number of components, $N$, selecting the appropriate model complexity based on the information content of the data.

We note that the underlying framework of \texttt{ThERESA}, \texttt{starry} \citep{starry}, takes the stellar and planetary masses and radii as input to set up the star-planet system for light-curve generation. We found discrepancies between light curves generated using \texttt{BATMAN} versus \texttt{starry} when holding the astrophysical and systematic parameters fixed. This was attributed to the fact that \texttt{starry} does not take a value of the semimajor axis over stellar radius, $a/R_{*}$ as input, but rather derives its own value from Kepler's third law using the inputted values of stellar mass, $M_{*}$, and orbital period, $P$. Using the \citet{southworth2012homogeneous} value for $M_{*}$ of 1.286 $\mathrm{M_{\odot}}$ gives a discrepant $a/R_{*}$ of 6.960, compared to the value of 7.110 we adopt from \citet{miritransit} in our \texttt{BATMAN} fits. For consistency in the orbital parameters between the fitting routines, we therefore re-derived a stellar mass value of 1.370 $M_{*}$ from Kepler's third law, using our adopted $a/R_{*}$ and $P$ from Table \ref{tab:system_and_fitting_params}. Using this value in the \texttt{starry} fits brings the light curves into agreement. We quote any additional parameters used in the \texttt{ThERESA} fit in Table \ref{tab:theresa_table}. All other fixed parameters in the fit are the same as those quoted in Table \ref{tab:system_and_fitting_params}.

\subsection{Model Selection} \label{sec:theresa_model_selection}

\begin{figure*}[hbt!]
\includegraphics[width=0.5\textwidth]{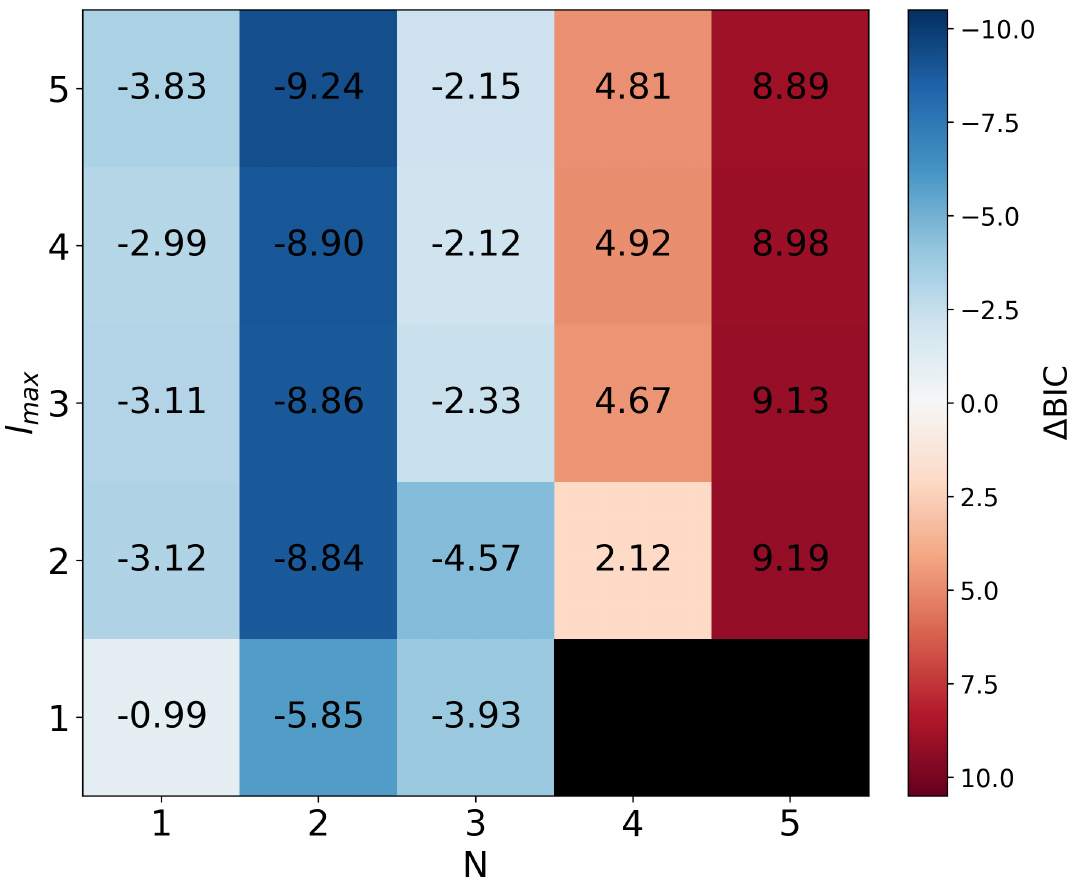}
\includegraphics[width=0.5\textwidth]{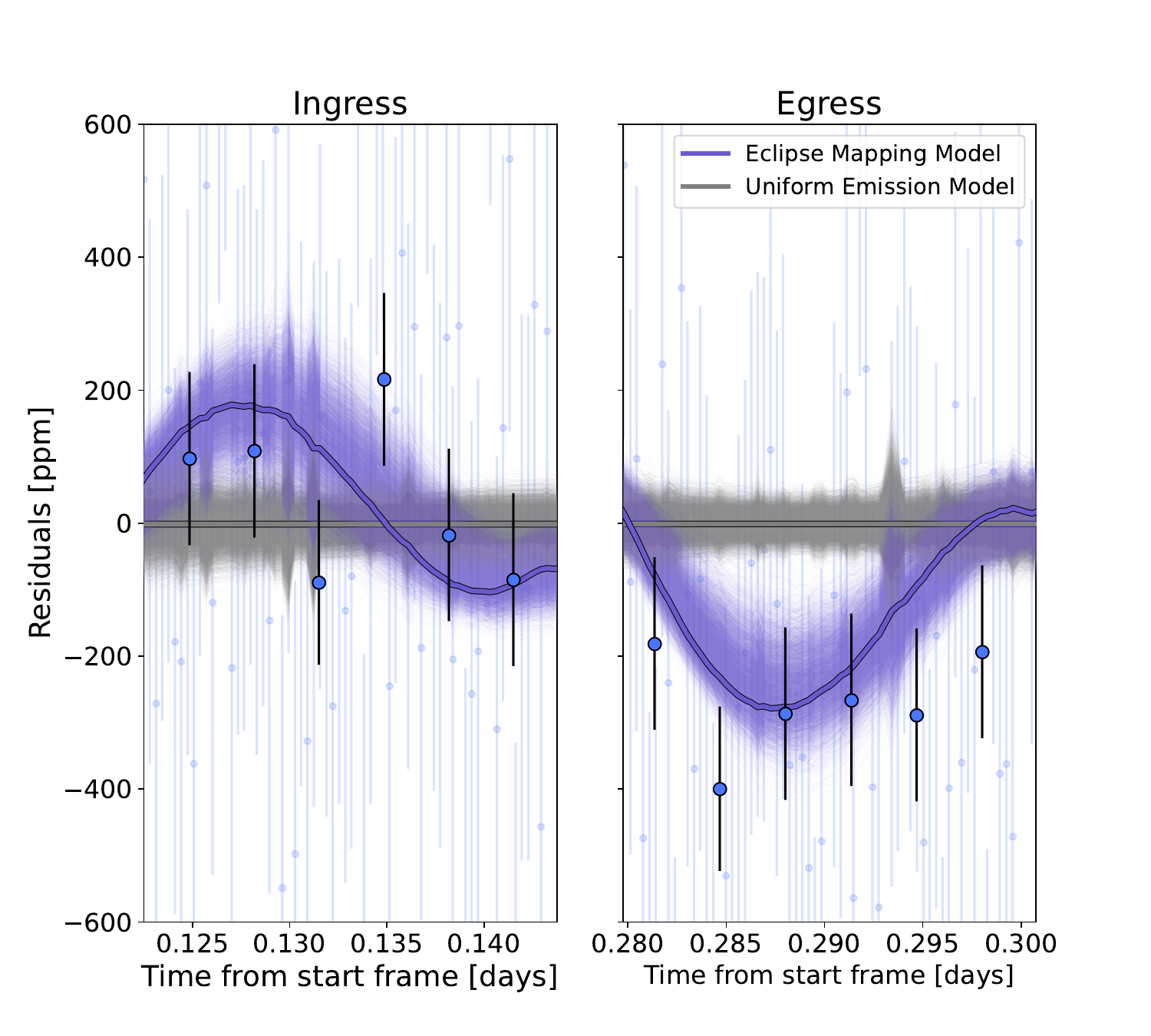}
\caption{(\textbf{Left}) Difference in BIC scores, $\Delta$BIC, between the fit to the data of \texttt{ThERESA} eclipse mapping models of varying maximum degree, $l_{max}$, and number of basis components, $N$, versus that of a uniform spatial emission model. A positive score indicates that the eclipse mapping model scores worse than the uniform model, whilst a negative score indicates that the eclipse mapping model scores better. \textbf{(Right)} Residuals of the uniform spatial emission model fit to the white light curve (blue data points, at native cadence in the background and 5 minute bins in the foreground). The grey lines are the range of uniform model fits, and the purple lines are the range of L2N2 eclipse mapping model fits post-processed from the posterior distribution, with the best-fit shown by the lines outlined in black. Residual deviations to the data in ingress/egress, which the uniform model is unable to adequately fit, is the eclipse mapping signal, which the eclipse mapping model fits to in order to produce the map.}
\label{fig:model_selection}
\end{figure*}

Before attempting to fit to the eclipse mapping signals, we first fit a uniform spatial emission \texttt{ThERESA} model (i.e., no eclipse mapping components) to the uncalibrated white light curve using the same systematic model as in the \texttt{BATMAN} fit (Equation \ref{eq:exotic_miri_light_curve_systematics}) to ensure that the two fitting routines converge on consistent best-fit parameters. The purpose of this consistency check is so that differences in the fit when allowing for spatially non-uniform emission can then be directly attributed to the eclipse mapping signals. The systematic parameters were fit with normal priors centred on the best-fit values determined from the \texttt{BATMAN} fit, with $1\sigma$ widths set as the \texttt{BATMAN} $3 \sigma$ uncertainties in order to allow for proper exploration of the parameter space. We adopt the same system parameters as in the \texttt{BATMAN} fit, including the \texttt{ExoTiC-MIRI} centre of transit time from \citet{miritransit}. Because the process of PCA orthogonlisation requires that the light curves are precomputed prior to fitting, \texttt{ThERESA} does not fit for system parameters (beyond eclipse depth via a scaling factor), and thus does not fit for centre of eclipse time, but rather fixes it to occur at 0.5 phase for a zero-eccentricity planet, as we assume for WASP-17b. This assumption is reasonable because the system age \citep[$\sim$3 Gyr,][]{anderson2011} is much greater than the timescales for circularisation of the orbit \citep[$\sim$5 Myr,][]{anderson2010}. Hence, we can make no comparison of the value of this parameter between the fitting routines. For all other parameters, both astrophysical and systematic, we found the best-fit values between the two fitting routines to be better than $1 \sigma$ consistent (see Table \ref{tab:theresa_table}), meaning that the fits to the light curve are statistically identical. We therefore proceed to fit to the eclipse mapping signals.

In the eclipse mapping fits, the eigenmap components were fit using uniform priors bound between -0.1 and 0.1. Note that these priors are very wide; we expect the values of these parameters to be on the order of the planetary-to-stellar flux (i.e., at most a few thousand ppm). We fit both the calibrated and uncalibrated white light curve to test the effect of jointly fitting systematics with the eclipse mapping signal. In the uncalibrated fit, we again set the systematic parameter priors to be $3 \sigma$ that of the \texttt{BATMAN} uncertainty values, lest the inclusion of additionally fitting to the eclipse mapping signals have a non-negligible impact on the systematic fit. These priors are summarised in Table \ref{tab:theresa_table}. We found that it was important to fit to the uncalibrated light curve in order to accurately recover the eclipse mapping signal, because the smaller-scale signals (i.e., all those below the day-night contrast) can get divided out by the systematic model when fitting with a spatially uniform emission model, making it difficult to later recover the eclipse mapping signals from the calibrated light curve. Hence, we recommend fitting the systematics simultaneously in future eclipse mapping studies in order to accurately recover the signal.

In creating the map, because our basis maps are constructed from spherical harmonics, there are plausible solutions that are mathematically but not physically valid, namely regions of negative flux. We therefore enforce the physical constraint that the fluxes must be positive at all locations on the planet that are scanned by our data, which is achieved by heavily penalising the fit quality (via the chi-squared statistics) of those that do have negative flux regions. Per \citet{theresa}, we set no such constraints on regions of the planet that are not scanned during the observation ($|\phi|\gtrsim110^{\circ}$) because we have no data to validate our model here, and therefore do not wish to impose constraints that are simply a consequence of the continuity of spherical harmonics rather than justified by the data. Hence, we make no inferences on the profile of these regions. However, we note that none of the models show regions of negative flux in these regions, regardless.

\begin{figure*}[hbt!]
\centering
\includegraphics[width=0.95\textwidth]{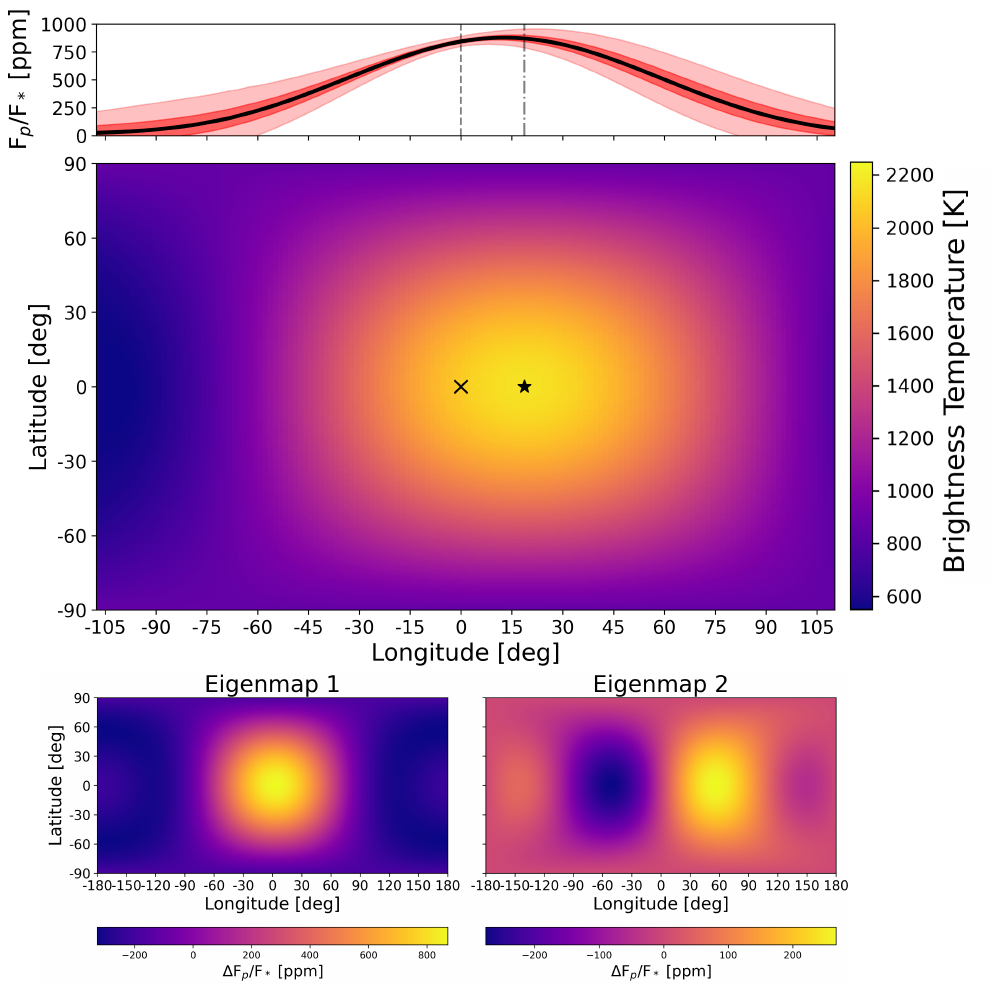}
\caption{\textbf{(Top)} Azimuthally-weighted zonal flux slice through the atmosphere, in units of eclipse depth, highlighting the decreasing flux contributions to the map going away from the substellar point and the increased uncertainties. (\textbf{Middle}) Broadband 5--12 $\upmu$m MIRI/LRS eclipse map of WASP-17b, in units of brightness temperature (see Section \ref{sec:map_components} for conversion details). The model shown is the L2N2 model. The cross symbol (and dashed line in the upper panel) marks the substellar point; the star symbol (and dash-dotted line in the upper panel) marks the hotspot. We omit the longitudes not scanned during the observation ($|\phi|\gtrsim110^{\circ}$) because we have no data to validate the model in these regions. \textbf{(Bottom)} Basis eigenmap components, scaled by their fitted weights. Values show the deviations that the maps incur from a spatially uniform map, in units of eclipse depth. The first eigenmap broadly corresponds to a day-night temperature contrast, and the second eigenmap to a hotspot offset.}
\label{fig:eclipsemap}
\end{figure*}

We tested an ensemble of eclipse mapping models up to a maximum spherical harmonic degree, $l_{max}=5$, and number of basis components, $N=5$ (resulting in a total of 24 models tested when accounting for the uniform model and the fact that for $l_{max}=1$, there are a maximum of $N=3$ basis components). We calculate the Bayesian Information Criterion \citep[BIC;][]{Raftery1995BIC} of each model in order to select the one that best captures the complexity of the data. The left panel of Figure \ref{fig:model_selection} shows the difference in BIC, $\Delta$BIC, between the \texttt{ThERESA} eclipse mapping models of varying $l_{max}$ and $N$ (hereafter referred to using the notation LxNx), compared to that of the aforementioned spatially uniform emission model fit (i.e., no eclipse mapping signal), also generated using \texttt{ThERESA}. A negative score indicates that that eclipse mapping model gives a statistically better fit to the data, whilst a positive score indicates that the uniform model gives a better fit. The better scores of the lower-order fits ($N\leq3$) compared to that of the uniform model indicates that there is a physical eclipse mapping signal in the data. The worse scores of the higher-order fits ($N\geq4$), on the other hand, tell us that these finer-scale signals are either not present in the data, or that we lack the precision to recover them. As such, it is the most observable large-scale features that the models are able to constrain. Namely, $N=2$ components are the most favoured, roughly corresponding to a day-night temperature contrast and a hotspot offset.

Whilst it is evident from this metric that $N=2$ models are statistically preferred, one can see that the appropriate $l_{max}$ is less clear, with all $N=2$ models with $l_{max} \geq 2$ scoring similarly well. The L5N2 model is the most statistically preferred, but only marginally over the other $N=2$ models with $l_{max} \geq 2$. Using

\begin{equation}
    P\left(\frac{M_{2}}{M_{1}} \right) = \mathrm{exp\left( -\frac{\Delta BIC}{2} \right)},
\label{eq:model_preference}
\end{equation}

\noindent to calculate the preference between two models \citep{Raftery1995BIC}, where $\mathrm{\Delta BIC=BIC_{2}-BIC_{1}}$ is the difference in Bayesian Information Criteria between models $M_{2}$ and $M_{1}$, we find that, with a $\mathrm{\Delta BIC=0.40}$, the highest-degree L5N2 model is only statistically preferred at a value 1.2$\times$ that of the lowest-degree L2N2 model.

The reason for the minor improvement of the higher-order fit, despite fitting for the same number of basis components, can be attributed to how the eigenmaps are constructed. In spherical harmonic mapping, increasing the degree to which one samples does not change the lower-degree basis maps, because the basis is fixed. In eigenmapping, on the other hand, the basis is not fixed, but in fact regenerated for each $l_{max}$ in order to maintain orthogonality between the eigencurves when including additional bases in the set. This regeneration of the basis as we increase $l_{max}$ may potentially non-negligibly alter the lower order eigenmaps in order for them to be additionally orthogonal to the newly introduced higher order eigenmaps.

In this case, it appears that the $N=2$ basis eigenmaps generated from higher-$l_{max}$ ensembles give marginally better fits to the data than those generated from lower-$l_{max}$ ensembles. The reason for their better BIC score despite the added complexity in the generation of these higher-$l_{max}$ models is that this step happens prior to the fitting. As far as the BIC metric is aware, these models fit using the same number of basis components and therefore are of an identical complexity; it is unable to account for the fact that a much larger ensemble of initial basis spherical harmonics was required in order to generate the higher-$l_{max}$ eigenmaps in the first place.

When inspecting the $N=2$ maps for $l_{max}\geq2$ models, we find highly consistent morphologies, with only minor differences in the longitudinal hotspot location that are better than 0.4$\sigma$ consistent between the maps. It is not possible to marginalise over these maps, because they are not constructed from the same functional set \citep{gibson2014, wakeford2016}. Given the similarity of these models, such efforts would also not meaningfully change the derived parameters nor their uncertainties. Therefore, given the only minor statistical preference for the L5N2 model over that of the L2N2 model, despite the significantly higher complexity involved in its generation, and the consistency between the resultant maps, we elect to use the simpler L2N2 model for subsequent analysis. We tabulate the best-fit parameters for this model in Table \ref{tab:theresa_table}.

In the right panel of Figure \ref{fig:model_selection}, we show the residuals of the uniform \texttt{ThERESA} model fit to ingress/egress. We show the data at its native cadence of 28 s in the background, and binned to $\sim$5 minute intervals in the foreground to increase the SNR and highlight the signal. In grey, we plot the range of uniform fits post-processed from 10,000 draws of the posterior distribution, with the best-fit model outlined in black. The binned data clearly deviate from this model, particularly in egress, where the range of feasible uniform model fits is unable to fit these excess residuals. In purple, we show the same post-processed distribution of light-curve fits for the L2N2 eclipse mapping model, which is able to achieve a much better fit to these excess residuals. This is the eclipse mapping signal from which we derive our map. With a $\Delta$BIC=8.84, this eclipse mapping model is preferred at a value 83$\times$ higher than that of the spatially uniform emission model, highlighting the clear detection of an eclipse mapping signal in this dataset.

\subsection{Map Components} \label{sec:map_components}

We show the broadband 5--12 $\upmu$m MIRI/LRS eclipse map of WASP-17b in the middle panel of Figure \ref{fig:eclipsemap}, corresponding to the L2N2 model, in units of brightness temperature. Because the map is generated from a spherical harmonic formulation, it naturally extends over the entire sphere of the planet. However, we elect to omit the longitudes of the planet that are not scanned during the observation ($|\phi|\gtrsim110^{\circ}$) because we have no data to validate the model in these regions. The conversion from flux to brightness temperature is performed using the same Phoenix model discussed in Section \ref{sec:reduction_comp}, and is additionally weighted by the MIRI/LRS throughput (although we note that this weighting has a negligible impact on the temperatures within their uncertainties). We do not account for uncertainties in the stellar parameters here because the uncertainties from the fit to the mapping signals will dominate, and also, per \citet{hammond2024}, in order to highlight the mapping uncertainties themselves.

The basis eigenmaps for the L2N2 model are shown in the bottom panel of Figure \ref{fig:eclipsemap}, scaled by their fitted weights. The colour scale indicates the flux deviations that the eigenmaps incur in the overall map from that of a spatially uniform emission model. The first eigenmap broadly corresponds to a day-night temperature contrast, which we recover at 3.8$\sigma$ significance. A maximum dayside brightness temperature of $2166 \pm 103$ K is found at the hotspot, with a latitudinally-averaged median dayside temperature of $1599 \pm 97$ K, in firm agreement with the blackbody temperature we derive from our emission spectrum in Section \ref{sec:reduction_comp}. This latitudinal averaging is performed by weighting the temperature profile by the squared cosine of the latitudes in order to account for both decreasing emitting area and reduced visibility due to viewing geometry at higher latitudes. Using the same weighting, we calculate a median western limb temperature of $625 \pm 231$ K and an eastern limb temperature of $1123 \pm 195$ K. Overall, these values give a day-night temperature contrast on the order of 1000 K, with limb differences of $\sim$500 K. These dayside temperatures, day-night temperature contrast, and limb temperature differences (within uncertainties) are in good agreement with the GCM results of \citet{kataria2016} for WASP-17b, which predict a maximum dayside temperature of $\sim$2200 K, a day-night temperature contrast of $\sim$1000 K, and limb differences of $\sim$300--400 K.

With regard to the reliability of these values, we refer to the upper panel of Figure \ref{fig:eclipsemap}, which shows the similarly weighted zonal flux contribution to the map. It is evident from this figure that the regions around the hotspot massively dominate the contribution to the map, with a value approximately 1000$\times$ that of the limb contributions. This is primarily due to the fact that the dayside is much hotter than the nightside and so is inherently brighter, but also partially because the dayside is observed continuously throughout the observation, whereas the limbs rotate out of view during the orbit and so are only briefly observed. The former factor is encapsulated by the drop-off in flux toward the limbs, and the latter by the large increase in uncertainty.
For these reasons, the model is primarily driven by the dayside profile between approximately -60 and 75$^{\circ}$ longitude, and so it is in these regions that the map is most reliable.

We also note that the model can only encapsulate uncertainties associated with the basis components it comprises. That is to say, because this is a low-degree model representing large-scale brightness features, it can only account for the associated large-scale uncertainties. At small scales below the resolution of these features, the uncertainties are unreliable. As such, only the large-scale temperature gradients are reliable (i.e., the day-night temperature contrast itself), whilst on small scales, the values are a byproduct of model continuity and largely not driven by the data (i.e., the limb-temperatures themselves). For a more thorough discussion of the nuances of brightness temperature uncertainties in eclipse maps, we refer the reader to \citet{theresa}.

The second eigenmap component, which we recover at 3.5$\sigma$ significance, corresponds to a longitudinal offset of the hotspot from the substellar point, which we recover to be $18.7^{+11.1\circ}_{-3.8}$ at the pressures probed by this data. The hotspot has a fairly sharp peak, and so its offset is the cause of the disparate limb temperatures because it concentrates significantly more of the dayside heat in the eastern half of the hemisphere than the western half. This hotspot offset indicates that the dominant mechanism for the transportation of heat from the substellar point does not occur via radial winds, which would result in a localised hotspot, but rather through the formation of a zonal equatorial jet \citep{lewis_hammond}. The relatively small value of the offset indicates that this redistribution of heat is somewhat inefficient, consistent with the value that we derive from our forward-model fits to the emission spectrum in Section \ref{sec:modelling_results}. These findings are in agreement with the GCM results of \citet{kataria2016}, which showed that high-temperature hot Jupiters like WASP-17b are expected to have hotspot offsets of $\sim$20--30$^{\circ}$ at pressures typically probed in emission.

We post-process from the posterior distribution of model light-curve fits the hotspot offset we would expect to measure via phase mapping, which recovers only large-scale features and so can only capture the hemispheric-average location of the hotspot. We find that, for the L2N2 model, we would recover a consistent longitudinal hotspot offset of $18.6^{+12.6\circ}_{-4.8}$ from phase mapping, with the marginally larger uncertainties attributed to the comparatively lower resolution of phase mapping compared to eclipse mapping. Given the fact that we only recover large-scale features in this eclipse map, we would expect to find such consistency. However, we find that the median hotspot location is shifted to marginally greater longitudes with larger upper bounds as we increase the $l_{max}$ of the $N=2$ models, to a maximum value of $23.4^{+16.4\circ}_{-4.4}$ for the L5N2 model, whilst the post-processed phase map value remains the same. These higher-degree basis maps recover smaller-scale structures that are only recoverable via the high-resolution scanning of eclipse mapping, and are inaccessible to the hemisphere-averaged inferences of phase mapping. This suggests that, despite only recovering large-scale features, there may be indications of small-scale longitudinal asymmetries to the hotspot structure present in the data, with more of the brightness skewed eastward than westward, which would be consistent with the chevron morphology theoretically expected for this feature \citep{showman2011}. This could potentially further explain the marginal preference for the L5N2 model over lower $l_{max}$, although we note that such small-scale structures are notoriously difficult to recover with accuracy due to their small-scale signals and spatial extent, which often place them in the eclipse mapping ``null space'' \citep{nullspace}. Hence, the statistical preference for these higher-degree models is not truly justified by the complexity of the data, with further observations being required in order to confirm such theories. Our eclipse mapping analysis of the NIRISS/SOSS and NIRSpec/G395H eclipses as part of this program may also potentially shed some light on this matter (D. Valentine et al. in preparation, see Section \ref{sec:conclusions} for details).

\subsection{Comparison to Eureka!}
For completeness, we perform the same eclipse mapping analysis on the \texttt{Eureka!} white light curve. Again, we adopt system parameters consistent with this reduction (see Table \ref{tab:system_and_fitting_params}), and scale additional parameters required for eclipse mapping accordingly, including the stellar mass in order to be consistent with the \texttt{BATMAN} light curves. We find similar results to those found for the \texttt{ExoTiC-MIRI} reduction, with a preference for low-order ($N\leq3$) models over a uniform model, heavy disfavouring of higher-order models, and highest preference for $N=2$ models. We also find the same plateau of $\Delta$BIC scores for all $N=2$ models with $l_{max}\geq2$, with consistent results found between the lowest- and highest-complexity models, L2N2 and L5N2, respectively. As such, we also determine the L2N2 model to best match the complexity of the data for the \texttt{Eureka!} reduction, and perform our comparison against this model. We find a larger day-night temperature contrast for the \texttt{Eureka!} eclipse map of $\sim$1500 K, which is greater than is to be expected for even the highest-temperature hot Jupiters \citep{kataria2016}. We make no comparison on the detection significance of the day-night contrast because such large-scale features are not localised to a single eigenmap, so comparing the values between maps produced from different datasets is not trivial. The hotspot offset, however, is associated purely with the second eigenmap, and can thus be compared between the models. We find consistent median values between the two reductions, but with tighter constraints on the \texttt{Eureka!} value of $15.0^{+3.9\circ}_{-2.8}$ and a higher detection significance of 4.3$\sigma$. However, the fit to the \texttt{ExoTiC-MIRI} reduction achieves a reduced chi-squared of 1.17 and a BIC of 1431.59, whereas the fit to the \texttt{Eureka!} reduction achieves much poorer values of 2.62 and 3119.45, respectively. Note that the BIC scores are essentially directly comparable because the light curves are identically trimmed and fit with the same number of free parameters in the model. 

To investigate the difference in fit quality, we can look to the error misestimation factor, $\beta$. The \texttt{BATMAN} fit to the \texttt{Eureka!} white light curve derives a $\beta$ of 1.63, indicative of underestimated flux uncertainties.
The tighter constraints on the hotspot offset from \texttt{Eureka!} are therefore likely unrepresentative of the true information content of the data. Comparatively, the $\beta$ value of 1.09 derived from the \texttt{BATMAN} fit to the \texttt{ExoTiC-MIRI} white light curve indicates that the derived flux uncertainties more accurately reflect the true data quality.

When scaling the flux uncertainties for both reductions by their respective $\beta$ values, we find consistent reduced chi-squareds of 1.00, with a BIC of 1222.45 for the \texttt{ExoTiC-MIRI} reduction and 1224.04 for the \texttt{Eureka!} reduction. We now find consistent day-night contrasts between the reductions of $\sim$1000 K. The uncertainties on the hotspot offset also increase, marginally for \texttt{ExoTiC-MIRI} to $18.7^{+12.2\circ}_{-4.3}$, and more significantly for \texttt{Eureka!} to $15.0^{+8.3\circ}_{-4.7}$. The upper bound of the \texttt{Eureka!} derived hotspot offset is still better constrained than that of the \texttt{ExoTiC-MIRI} value, but due to the more significant scaling of the flux uncertainties ($\beta$=1.63 compared to 1.09), the detection significance decreases to 2.6$\sigma$ compared to 3.2$\sigma$ for the \texttt{ExoTiC-MIRI} result. We interpret this as being a consequence of the \texttt{ExoTiC-MIRI} custom linearity correction: the higher detection significance is a result of the increased accuracy it grants, whilst the higher upper uncertainty is a result of the sacrifice in precision required in order to attain this higher accuracy.

This comparison demonstrates the importance of accurate derivation of the flux uncertainties for eclipse mapping purposes, and this will be important to take into consideration when utilising data from multiple instruments where detector gain uncertainties vary. These nuances will be explored further in D. Valentine et al. (in preparation) when combining eclipse mapping results across multiple JWST instruments. 

\subsection{Centre of Eclipse Timing Offset}
\begin{figure}[t]
\centering
\includegraphics[width=0.475\textwidth]{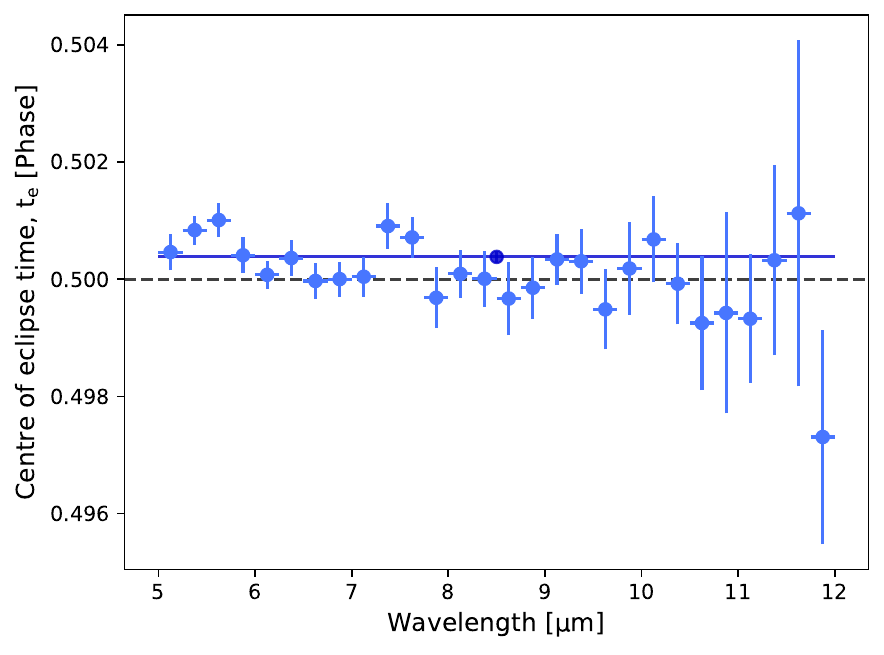}
\caption{Timings of central eclipse found in the \texttt{BATMAN} fits to the 0.25 $\upmu$m bin spectroscopic light curves (\texttt{ExoTiC-MIRI} reduction). Overlaid is the white light curve value, showing the broadband average.}
\label{fig:eclipse_timings}
\end{figure}

The longitudinal hotspot offset induces an apparent timing offset to the centre of eclipse time, altering it from the expected value of 0.5 phase for zero-eccentricity planets. We find consistent timing offsets between the \texttt{BATMAN} fits to the white light curve of each reduction of approximately 2 minutes (corresponding to $\sim$0.0004 phase, as shown in Figure \ref{fig:eclipse_timings}). To quantify whether the magnitude of this offset could equally be explained by an eccentric orbit, we use \texttt{BATMAN} to re-fit the white light curve with eccentricity as a free parameter. We recover consistent eccentricities between the reductions of $e=0.013\pm0.005$ for \texttt{ExoTiC-MIRI} and $e=0.011\pm0.005$ for \texttt{Eureka!}. The inclusion of this non-zero eccentricity in the fit reduces the timing offset by only approximately 10 seconds in both reductions. Given that the eccentricity of WASP-17b is tightly constrained to be $<$0.02 \citep{wasp17eccentricity}, an eccentric orbit is therefore insufficient to explain the magnitude of our observed offset. Instead, we attribute it to the non-uniform brightness pattern of the dayside. The longitudinal shift of the hotspot results in one half of the hemisphere being brighter than the other. As such, the point at which half of the brightness of the dayside hemisphere has been eclipsed is not coincident with the point at which half of the hemisphere has been physically obscured, as is assumed in a model fit where we assume the emission to be spatially homogeneous, which is the case for \texttt{BATMAN}. In the case of WASP-17b, the eastward shift of the hotspot means that the eastern hemisphere is overall brighter than the western hemisphere; therefore, the centre of eclipse time occurs later than expected.

In our spectroscopic fitting, we find a range of centre of eclipse timing offsets, from zero to a maximum of approximately 5 minutes (excluding the anomalous and less reliable final channel), as shown in Figure \ref{fig:eclipse_timings} (note that 0.001 phase $\sim$ 5.5 minutes). This indicates that there is a range of atmospheric dynamics across the pressure space probed by this dataset, the bulk properties of which are captured within this broadband eclipse map. Groupings of timing offsets give indications of which wavelengths probe similar pressures in the atmosphere (e.g. 5--6, 7--8 $\upmu$m), and can therefore be used to guide binning regimes for spectroscopic eclipse mapping \citep{theresa}. Such efforts will be employed in D. Valentine et al. (in preparation).

\subsection{Latitudinal Constraints}
In principle, one of the advantages of eclipse mapping over phase mapping is that we are able to map the latitudinal profiles of exoplanet atmospheres. However, this signal is only present during eclipse ingress/egress, whereas the longitudinal signal is also present in the eclipse baseline as the planet rotates. Additionally, the detectability of the latitudinal signal depends strongly on the impact parameter of the planet. By considering eclipse geometry, is is evident that, in order to achieve the finest slicing over the latitudes (which corresponds to higher spatial resolution and more detectable signals), we require a planet with a high impact parameter. We therefore theoretically achieve the best longitudinal signal the lower the impact parameter, and conversely the best latitudinal signal the higher the impact parameter, as these configurations give the finest sampling over the respective dimensions. However, the fact that the longitudinal signal is also present out-of-eclipse increases its inherent detectability over the latitudinal signal. Hence, in order to achieve the optimal balance of both longitude and latitude signal detectability, it is best to have a higher impact parameter in order to boost the latitudinal signal strength in ingress/egress. Additionally, higher impact parameters result in longer ingress/egress durations, which increases the eclipse mapping signal as a whole. As advocated by \citet{boone2024}, an impact parameter of approximately 0.7 best balances these factors to give the optimal geometric eclipse mapping metric for a given planet. The 0.35 impact parameter of WASP-17b is sub-optimal for recovering latitudinal signal, hence our inability to constrain the latitudinal profile from this single eclipse.

Given high enough precision, however, the eclipse geometry of WASP-17b could potentially be sufficient to constrain the latitudinal profile. The left panel of Figure \ref{fig:model_selection} shows that $N=3$ models, which include a third basis eigenmap with latitudinal asymmetry (the third most observable eclipse mapping signal) are still statistically preferred over that of a uniform hemisphere fit, indicating that there is potential evidence of latitudinal signals in the data. We inspected these $N=3$ maps and found temperature gradients consistent with those of the $N=2$ maps. The inclusion of a latitudinal offset to the hotspot results in a larger distribution of plausible longitude-latitude values in the posterior, including some unphysical solutions that place the hotspot on the nightside. Nonetheless, there is a peak in the parameter space that gives a consistent longitudinal hotspot offset to the $N=2$ models of $\sim$20$^{\circ}$, and a latitudinal hotspot offset of $\sim$30$^{\circ}$.

Latitudinal hotspot offsets can be induced by marginal misalignments between the spin and magnetic field axis of an exoplanet, which can perturb the equatorial jet from its latitudinally symmetric state via Ohmic dissipation \citep{rogers2014}. Hot Jupiters with temperatures of 1400--2000 K are most susceptible to this process, and so it is feasible that we might observe this phenomenon for WASP-17b. With regard to the probability of a spin-dipole misalignment, it is challenging to constrain this value for unresolved transiting exoplanets, but such misalignments are observed for the majority of the Solar System gas giants \citep{stevenson2003}, and so it is plausible that we might observe the same misalignment for a number of exoplanets. The MIRI/LRS eclipse map of WASP-43b, for example, also shows evidence of latitudinal asymmetries in its brightness pattern \citep{hammond2024}, which could potentially also be attributed to such effects. Ohmic dissipation in the deep convective regions of a planet has also been proposed as a mechanism for radius inflation \citep{ohmic_inflation}, which would be consistent with the highly inflated radius of WASP-17b. This coupling between the flow and magnetic field deep in the interior would also result in elevated internal temperatures, which is consistent with the proposed explanation for excess flux observed at short wavelengths in the NIRISS/SOSS emission spectrum of WASP-17b (A. Gressier et al. accepted, AJ). Additional observations to increase our precision would be required in order to potentially better constrain the latitudinal profile and explore the possibility of latitudinal asymmetries to the spatial emission profile of WASP-17b, although again, our eclipse mapping analysis of the NIRSPEC/G395H and NIRISS/SOSS datasets (D. Valentine et al. in preparation) may also shed light on the matter.

\subsection{Comparison to Existing JWST Eclipse Maps}
Comparing our map to other existing JWST eclipse maps, \citet{wasp18eclipsemap} find no such longitudinal hotspot offset in their eclipse map of WASP-18b. Ultra-hot Jupiters are not predicted to form appreciable equatorial jets because, at these temperatures, the radiative timescales become much shorter than the timescales for the propagation of Kelvin and Rossby waves \citep{kataria2016}. Radial winds from the substellar point dominate the heat redistribution at these temperatures, which is less efficient than through the means of an equatorial jet, leading to even more extreme day-night temperature contrasts in excess of 1500 K. These high dayside temperatures are also sufficient to ionise trace alkalies in the atmosphere, which can couple to the planetary magnetic field and induce Rayleigh drag, which serves to further localise the hotspot \citep{beltz2022}.

\citet{hammond2024} find a $7.75\pm0.36^{\circ}$ longitudinal hotspot offset in their eclipse map of WASP-43b, with these tight constraints largely driven by the long-baseline phase signal. Despite its colder equilibrium temperature of $\sim$1400 K, GCM simulations of WASP-43b predict it to have a similar longitudinal hotspot offset to WASP-17b of $\sim$20--30$^{\circ}$ as it is a faster rotator, with a period of 0.8 days compared to 3.7 days for WASP-17b. However, enhanced metallicities can also seek to reduce longitudinal hotspot offsets, with a 5$\times$Solar enhancement for WASP-43b consistent with a smaller hotspot offset in the region of 10--20$^{\circ}$ \citep{katariawasp43}. Previous HST/WFC3 dayside emission measurements of WASP-43b are consistent with such enhanced metallicities \citep{stevenson2014, kreidberg2015batman}, and the smaller hotspot offset derived from the eclipse map is also in general agreement with previous values in the region of 10$^{\circ}$ inferred from shifts of the peak phase amplitude \citep{stevenson2014}. Our program's previous work found only a slightly supersolar metallicity for WASP-17b \citep[][A. Gressier et al. accepted, AJ]{miritransit}, similarly to the median of our forward-model fits (despite our larger uncertainties). These values are therefore consistent with our eclipse mapping findings of no significant reduction to the longitudinal hotspot offset. However, we note that, due to our looser constraints on the hotspot location compared to those of \citet{hammond2024}, we cannot make comparably strong inferences on parameters which may be altering the hotspot offset from its expected value, which include not only metallicity, but also the likes of magnetic field effects, atmospheric composition, and frictional drag.

In addition, \citet{hammond2024} discuss the possibility that their value is an underestimation of the true offset because their map, similarly to ours for WASP-17b, is low-order and so only comprises large-scale features. The omission of small-scale structure may lead to misestimation of the hotspot location, and this too may explain our differing median hotspot location between lower-order (L2N2) and higher-order (L5N2) models for WASP-17b. However, the effect is more severe for cooler planets like WASP-43b where the day-night temperature contrast and therefore thermal homogenisation is smaller, leading to a less uniform profile with a higher degree of small-scale structure. Our larger uncertainties on the hotspot location also mean that the impact of such inaccuracies is further minimised in our case, whereas the tight constraints of \citet{hammond2024} further exacerbate the problem. Hence, while these small uncertainties are accurate to the data quality of the map, they may be inaccurate to the true profile of the planet. This highlights the potential biases in using observable hotspot offsets to make inferences on the atmospheric dynamics, and the care that must be taken in doing so. In particular, it is likely that observable hotspot offsets may often be underestimations of the ``true" value due to the omission of the chevron morphology expected for the hotspots of hot Jupiters with offsets induced via equatorial jets \citep{showman2011}, as has been shown theoretically in numerous instances \citep{nullspace, hammond_cv}. For WASP-17b, our larger upper uncertainty on the hotspot location reflects this potential for underestimation, giving us greater reliability in our results.

\section{Conclusions \& Future Work} \label{sec:conclusions}
We have presented the first mid-IR spectroscopic characterisation of the dayside atmosphere of WASP-17b using a single MIRI/LRS eclipse from 5--12 $\upmu$m. We performed two independent reductions using the \texttt{ExoTiC-MIRI} and \texttt{Eureka!} pipelines, which yielded highly consistent emission spectra, demonstrating the robustness of our reduction methodologies. We found our choice of background subtraction to have the most significant impact on our derived emission spectra. We recommend a row-by-row treatment to adequately model the wavelength-dependent profile, and advise exploration of the appropriate-order polynomial to best match the complexity of the data. Similarly to what was found in the analysis of the MIRI/LRS transit of WASP-17b presented by \citet{miritransit}, we also found our choice of linearity correction to have an impact on our derived stellar spectra. Through inspection of the group-level ramps, we found that deriving a self-calibrated linearity correction from only a subset of the most ``well-behaved" groups, and removing groups which are visibly affected by systematics \citep[e.g.,][]{ressler2015mid, argyriou2023brighter, wright2023mid}, may lead to more reliable results. However, this increase in accuracy comes at the cost of reduced precision as opposed to utilising the entirety of the groups for the ramp-fitting.

\texttt{PICASO} forward-model fits \citep{batalha_picaso} to the spectra identified the presence of water in the dayside atmosphere, consistent with findings from the NIRISS/SOSS emission spectrum of WASP-17b (A. Gressier et al. accepted, AJ). We also identify the potential signatures of unidentified spectral features, which may be attributed either to white noise/unknown systematics, or potentially as yet unidentified cloud species (e.g., see Figure \ref{fig:clouds}). We performed a Gaussian significance test on these features in order to quantify their detection significance, yielding values of 2.8$\sigma$ and 2.5$\sigma$ for the \texttt{ExoTiC-MIRI} and \texttt{Eureka!} reductions, respectively. These values give good indications of a likely astrophysical origin, but our limited cloud analysis of these features was unable to properly model them, which is unsurprising given the limited chemical information content of MIRI/LRS emission data for high-temperature hot Jupiters. We expect that our analysis of the 0.6--12 $\upmu$m emission spectrum of WASP-17b from individual eclipses observed using JWST NIRISS/SOSS, NIRSpec G395H, and MIRI/LRS may elucidate such theories, and we will devote a more comprehensive analysis to that work (H.R. Wakeford et al. in preparation). 

We found that this MIRI/LRS emission spectrum was unable to adequately constrain three of the four parameters in our model grid, those being internal temperature, metallicity, and C/O ratio. The latter two results can be attributed to the aforementioned limited chemical content of the data, and the former to the low pressures it probes where interior effects are less significant. However, this bandpass is rich in thermodynamical information content, and so our model fits were able to derive a tightly constrained value of the heat redistribution factor, with a median value of $0.92\pm0.02$ between the reductions, indicative of inefficient heat redistribution from the dayside to the nightside. This is in agreement with results derived from the MIRI/LRS transmission spectrum of WASP-17b \citep{miritransit}, and additionally consistent with the efficient heat redistribution derived from the NIRISS/SOSS emission spectrum (A. Gressier et al. accepted, AJ) when accounting for the fact that they probe higher pressures in the atmosphere where the thermal profile becomes homogenised by interior contributions, whereas our data probes lower pressures where stellar irradiation of the substellar point primarily dominates the thermal profile, producing a starker day-night temperature contrast \citep{kataria2016}.

Due to the rich thermodynamical content of this MIRI/LRS data, we utilised the eclipse mapping software \texttt{ThERESA} to extract the signals of WASP-17b's dayside photospheric spatial emission profile imprinted on the eclipse light curve during ingress/egress. We found that fitting the systematics simultaneously is important in order to properly extract the smaller-scale eclipse mapping signals. This is because the systematic model may attempt to additionally fit to these signals in a uniform model fit since they are inconsistent with the assumptions of the astrophysical model (i.e., smoothly decreasing flux in ingress/egress). Namely, we find that the linear detrending against time (parameterised by $c_{0}$ in the systematic model) seeks to divide out the asymmetry induced by the eastward hotspot offset for this dataset, which results in a surplus of flux during egress and dearth during ingress due to the asymmetry of the hemisphere's brightness pattern. This is reflected by the altered values between the uniform and eclipse mapping model fits in Table \ref{tab:theresa_table}.

From our eclipse mapping fits to the white light curve, we find that an L2N2 model best represents the complexity of the data, and we use this model to construct the first eclipse map of WASP-17b. This broadband 5--12 $\upmu$m map comprises two basis components: a day-night temperature contrast of order 1000 K and a longitudinal hotspot offset of $18.7^{+11.1\circ}_{-3.8}$. These features are indicative of the presence of an equatorial jet transporting stellar-irradiated heat zonally from the substellar point and dominating the global heat recirculation, in agreement with theoretical expectations for hot Jupiters \citep{showman2011}. The fairly sharply peaked morphology and magnitude of the hotspot offset is in agreement with the 20--30$^{\circ}$ GCM prediction for high-temperature hot Jupiters, including WASP-17b, as are the resultant limb temperature difference of a few hundreds of kelvin and large day-night temperature contrast \citep{kataria2016}. These indicate that the jet's recirculation of heat from the substellar point is inefficient, which is consistent with the poor heat redistribution factor that we derive from our forward-model fits to the emission spectrum. There is potential evidence for an asymmetry to the shape of the hotspot, namely skewing it east, consistent with the expected chevron morphology of this feature \citep{showman2011}. There are also indications of a potential latitudinal offset to the hotspot; such effects have been theorised \citep[e.g.,][]{batygin2014, komacek2020, ch02021}, but not adequately explored either theoretically or observationally, primarily as a result of our inability to constrain the latitudinal profile of transiting exoplanets prior to JWST. However, we found that the complexity of the data was not sufficient to adequately constrain either of these potential features. Further observations would be required in order to increase the precision of the data, particularly in ingress/egress, which is the only region of the light curve where the latitudes are scanned.

This eclipse mapping analysis of WASP-17b will be expanded on in D. Valentine et al. (in preparation). The aims of the JWST-TST DREAMS program is to carry out comprehensive multidimensional characterisation of three transiting exoplanets representative of key exoplanet classes: Hot Jupiters (WASP-17b, GTO~1353), Warm Neptunes (HAT-P-26b, GTO~1312), and Temperate Terrestrials (TRAPPIST-1e, GTO~1331). As part of the GTO~1353 program, we have observed WASP-17b in eclipse using JWST NIRISS/SOSS (A. Gressier et al. accepted, AJ), NIRSpec/395H (H.R. Wakeford et al. in preparation), and MIRI/LRS (this work). These observations span 0.6--12 $\upmu$m, and therefore probe a wide range of pressures in the atmosphere. Additionally, owing to the high precision of JWST light curves, it is feasible to achieve high enough SNR to eclipse map spectroscopic light curves for single eclipse observations,
which allows us to sample more distinct pressure levels in the atmosphere, increasing our vertical resolution. This allows us to eclipse map the dayside atmosphere not only in longitude-latitude, but also in altitude, yielding three-dimensional eclipse maps. Per our program aims, this will grant us far greater insight into the multidimensional properties of exoplanet atmospheres, information that is imperative in order to properly interpret the wealth of information that JWST light curves provide.

\vspace{0.7cm}
We thank the anonymous referee for their comments that helped improve the quality of this paper. D.V. would like to thank M. Hammond for useful discussions on the techniques used in this paper. This paper reports work carried out in the context of the JWST Telescope Scientist Team (\url{https://www.stsci.edu/~marel/jwsttelsciteam.html}, PI: M. Mountain). Funding is provided to the team by NASA through grant 80NSSC20K0586. Based on observations with the NASA/ESA/CSA JWST, associated with program(s) GTO-1353 (PI: N.K. Lewis), obtained at the Space Telescope Science Institute, which is operated by AURA, Inc., under NASA contract NAS 5-03127. The JWST data presented in this paper were obtained from the Mikulski Archive for Space Telescopes (MAST) at the Space Telescope Science Institute. The specific observations analyzed can be accessed via \href{http://dx.doi.org/10.17909/cqsf-sx69}{10.17909/cqsf-sx69}, and data products and models are available at \url{https://doi.org/10.5281/zenodo.12571830}.

D.V. acknowledges funding from STFC grant ST/X508263/1 and from the University of Bristol School of Physics PhD Scholarship Fund. H.R.W. and D.G were funded by UK Research and Innovation (UKRI) framework under the UK government’s Horizon Europe funding guarantee for an ERC Starter Grant [grant number EP/Y006313/1].

\vspace{5mm}
\facilities{JWST(MIRI/LRS)}

\software{
\texttt{ExoTiC-MIRI} \citep{grantexoticmiri},
\texttt{Eureka!} \citep{eureka},
\texttt{ThERESA} \citep{theresa},
\texttt{BATMAN} \citep{kreidberg2015batman},
\texttt{emcee} \citep{foreman2013emcee},
\texttt{PICASO} \citep{batalha_picaso},
\texttt{Virga} \citep{batalha_virga},
\texttt{starry} \citep{starry},
\texttt{MC3} \citep{mc3},
\texttt{dynesty} \citep{dynesty},
\texttt{POSEIDON} \citep{poseidon},
\texttt{numpy} \citep{harris2020array},
\texttt{SciPy} \citep{2020SciPy-NMeth},
\texttt{matplotlib} \citep{Hunter:2007},
\texttt{xarray} \citep{hoyer2017xarray, hoyer_stephan_2022_6323468},
\texttt{astropy} \citep{astropy:2013, astropy:2018, astropy:2022}.
}

\appendix

\begin{deluxetable*}{l c c c c}[hbt!]
    \renewcommand{\arraystretch}{1.1}
    \tabletypesize{\footnotesize}
    \tablecolumns{5} 
    \tablecaption{\texttt{BATMAN} light-curve fitting parameter information. Values are shown for the white-light-curve fits.}
    \tablehead{ & \multicolumn{2}{c}{ExoTiC-MIRI} & \multicolumn{2}{c}{Eureka!}}
    \startdata
    Parameter & Prior & Value & Prior & Value \\
    \hline
    $P$ [days] & fixed \citep{Alderson2022} & $3.73548546$ & fixed \citep{Alderson2022} & $3.73548546$ \\
    $a/R_{*}$ & fixed \citep{miritransit} & $7.110$ & fixed \citep{sedaghati2016potassium} & 7.025 \\
    $i$ [degrees] & fixed \citep{miritransit} & $87.217$ & fixed\citep{sedaghati2016potassium} & 86.9 \\
    $\mathrm{R_p}/\mathrm{R_*}$ & fixed \citep{miritransit} & 0.12472 & fixed \citep{miritransit} &  0.12506\\
    $t_0$ [$\mathrm{BJD}_\mathrm{TDB} - 2400000$] & fixed \citep{miritransit} & 60016.726452 & fixed \citep{miritransit} &  60016.726477\\
    $F_P / F_*$ & $\mathcal{U}$(0, 0.01) & $0.002579 \pm 0.000056$ & $\mathcal{U}$(0, 0.05) & $0.002595 \pm 0.000047$ \\
    $t_e$ [$\mathrm{BJD}_\mathrm{TDB} - 2400000$] & $\mathcal{N}$(60018.594195, 0.01) & $60018.59562 \pm 0.00034$ & $\mathcal{N}$(60018.594220, 0.05) & $60018.59539 \pm 0.00032$ \\
    \hline
    $r_0$ & $\mathcal{N}$(0, 0.1) & $-0.00366 \pm 0.00018$ & $\mathcal{N}$(0, 0.01) & $-0.00385 \pm 0.00028$ \\
    $r_1$ & $\mathcal{U}$(0, 200) & $19.1 \pm 2.1$ & $\mathcal{U}$(5, 150)  & $20.2 \pm 1.7$ \\
    $c_0$ & $\mathcal{N}$(0, 0.01) & $-0.00269 \pm 0.00063$ & $\mathcal{N}$(0, 0.01) & $-0.00180 \pm 0.00049$ \\
    $s_0$ & $\mathcal{U}$(-250, 250) & $-7.60 \pm 0.92$ & ... & ... \\
    $f_0$ & $\mathcal{N}$(1.0, 0.01) & $1.000023 \pm 0.000047$ & $\mathcal{N}$(1.0, 0.01) & $0.998927 \pm 0.000050$ \\
    $x_0$ & ... & ... & $\mathcal{N}$(0, 0.1) & $0.0442 \pm 0.0065$\\
    $x_1$ & ... & ... & $\mathcal{N}$(0, 0.5) & $0.018 \pm 0.021$\\
    $\beta$ & $\mathcal{U}$(0.1, 5) & $1.09 \pm 0.02$ & $\mathcal{N}$(1.5, 0.5) &  $1.626 \pm 0.033$ \\
    \enddata
    \label{tab:system_and_fitting_params}
    \tablecomments{Parameter definitions: eclipse depth, $F_P / F_*$; time of eclipse centre, $t_e$; systematic model parameters as defined in Equation \ref{eq:exotic_miri_light_curve_systematics}, $r_0$, $r_1$, $c_0$, and $s_0$; baseline, $f_0$; decorrelation coefficients for trace position and width, $x_0$ and $x_1$; and error multiplier, $\beta$.}
\end{deluxetable*}

\begin{deluxetable*}{l c c c}[hbt!]
    \centering
    \renewcommand{\arraystretch}{1.25}
    \tabletypesize{\footnotesize}
    \tablecolumns{5} 
    \tablecaption{\texttt{ThERESA} light-curve fitting parameter information.}
    \tablehead{ & & L2N2 Model & Uniform Model}
    \startdata
    Parameter & Prior & Value & Value\\
    \hline
    $M_{*}$ [$\mathrm{M_{\odot}}$] & fixed$^*$ & \multicolumn{2}{c}{1.370} \\
    $R_{*}$ [$\mathrm{R_{\odot}}$] & fixed \citep{southworth2012homogeneous} & \multicolumn{2}{c}{1.583}\\
    $M_{p}$ [$\mathrm{M_{J}}$] & fixed \citep{southworth2012homogeneous} & \multicolumn{2}{c}{0.477} \\
    $R_{p}$ [$\mathrm{R_{J}}$] & fixed** & \multicolumn{2}{c}{1.921}\\
    $C_{0}$ & $\mathcal{U}$(-0.1, 0.1) & $0.00156^{+0.00044}_{-0.00030}$ & $0.002592^{+0.00046}_{-0.00045}$\\
    $C_{1}$ & $\mathcal{U}$(-0.1, 0.1) & $0.00059^{+0.00016}_{-0.00024}$ & /\\
    $C_{2}$ & $\mathcal{U}$(-0.1, 0.1) & $-0.000275\pm0.000076$ & /\\
    $s_{corr}$ & $\mathcal{U}$(-0.1, 0.1) & $0.00038^{+0.00015}_{-0.00013}$ & $0.00064^{+0.00014}_{-0.00012}$\\
    $r_{0}$ & $\mathcal{N}$(-0.00366, 0.00054) & $-0.00365^{+0.00015}_{-0.00014}$ & $-0.00376^{+0.00015}_{-0.00014}$\\
    $r_{1}$ & $\mathcal{N}$(19.1, 6.3) & $19.1^{+1.7}_{-1.6}$ & $18.8^{+1.6}_{-1.5}$\\
    $c_{0}$ & $\mathcal{N}$(-0.00269, 0.00189) & $-0.00174^{+0.00055}_{-0.00052}$ & $-0.00300^{+0.00049}_{-0.00045}$\\
    $s_{0}$ & $\mathcal{N}$(-7.60, 2.76) & $-7.60^{+0.81}_{-0.82}$ & $-7.72^{+0.82}_{-0.81}$\\
    \hline
    \enddata
\label{tab:theresa_table}
    \tablecomments{Parameter definitions:
    stellar and planetary masses and radii, $M_{*}$, $R_{*}$ and $M_{p}$, $R_{p}$, respectively;
    eigencurve fit weights, $C_{0}-C_{2}$; stellar baseline correction, $s_{corr}$; and systematic model parameters, $r_0$, $r_1$, $c_0$, and $s_0$, as defined in Equation \ref{eq:exotic_miri_light_curve_systematics}. Any additional fixed parameters in the fit are the same as those adopted in Table \ref{tab:system_and_fitting_params}.
    (*) Calculated using Kepler's third law with $a/R_{*}$ and orbital period, P, values adopted from Table \ref{tab:system_and_fitting_params}. See Section \ref{sec: theresa_method} for further information.
    (**) $R_{p}/R_{*}$ value from \citet{miritransit} scaled by $R_{*}$ value from \citet{southworth2012homogeneous}.}
\end{deluxetable*}

\clearpage
\onecolumngrid
\renewcommand\thefigure{\thesection\arabic{figure}}    
\section{Unexplained Spectral Features}
The effective cross sections of plausible cloud species that may potentially explain the spectral features at long wavelengths in the MIRI/LRS emission spectrum of WASP-17b are shown in Figure \ref{fig:clouds}. These are generated using POSEIDON \citep{poseidon} with Mie scattering (E. Mullens et al. in preparation). We perform a Gaussian significance test on these features to quantify the probability of them being of astrophysical origin, the results of which are presented in Figure \ref{fig:gauss_sig_test}.

\setcounter{figure}{0} 
\begin{figure*}[hbt!]
\includegraphics[width=0.99\textwidth]{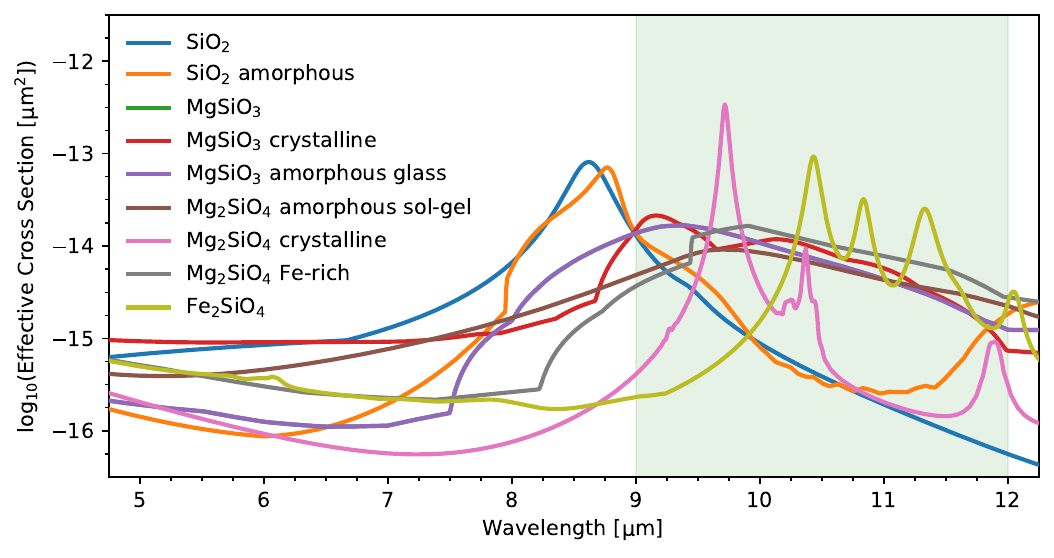}
\caption{Cloud condensate effective cross sections in the MIRI/LRS bandpass of plausible cloud species at the dayside temperatures of WASP-17b. The highlighted region indicates the wavelengths where we observe potential astrophysical features in the data, over which many of these species show prominent absorption features of a similar morphology to the features.}
\label{fig:clouds}
\end{figure*}

\begin{figure*}[hbt!]
\includegraphics[width=0.99\textwidth]{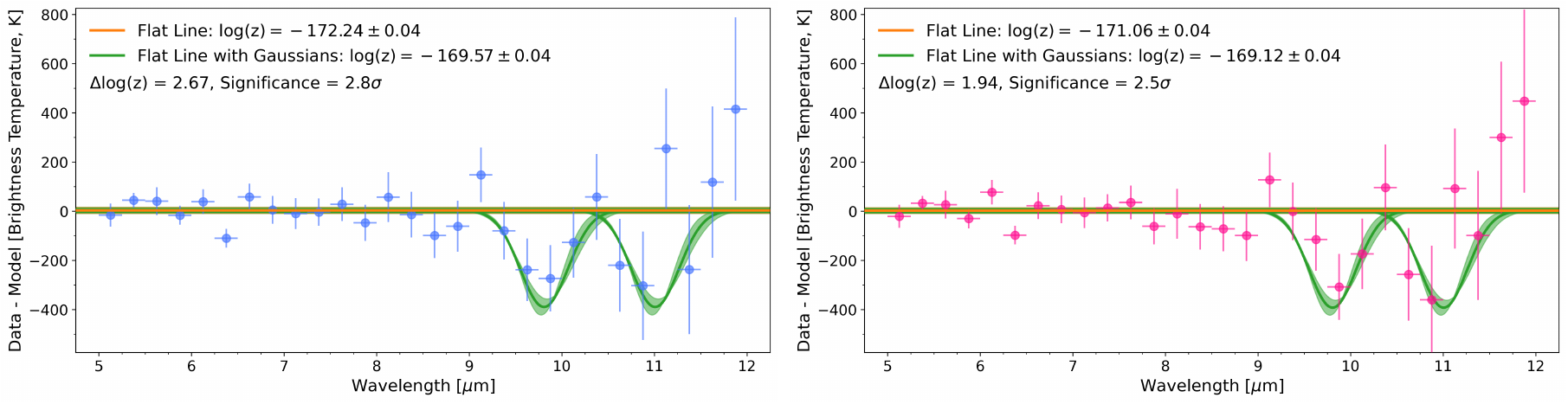}
\caption{Gaussian significance test on the unidentified spectral features in both the \texttt{ExoTiC-MIRI} (left) and \texttt{Eureka!} (right) spectra between 9 and 12 $\upmu$m. The data have the \texttt{PICASO} forward models presented in Figure \ref{fig:forward_model} subtracted in order to fit for deviations from the modelled physics. The orange line is a flat linear fit, which would be consistent with no additional features, while the green line incorporates Gaussian fits to potentially unmodelled spectral features.}
\label{fig:gauss_sig_test}
\end{figure*}

\clearpage

\bibliography{WASP17_JWST_TST_MIRI_ECLIPSE}{}
\bibliographystyle{aasjournal}

\end{document}